\documentclass[fleqn,usenatbib]{mnras}

\usepackage[utf8]{inputenc} % set input encoding (not needed with XeLaTeX)
\usepackage{times} 
\usepackage[T1]{fontenc} 
\usepackage{aecompl}
\usepackage{newtxtext,newtxmath}
\usepackage{amsmath}
\usepackage[T1]{fontenc}

\DeclareRobustCommand{\VAN}[3]{#2}
\let\VANthebibliography\thebibliography
\def\thebibliography{\DeclareRobustCommand{\VAN}[3]{##3}\VANthebibliography}

\usepackage{geometry} % to change the page dimensions
\usepackage{graphicx,url} % support the \includegraphics command and options

\usepackage{booktabs} % for much better looking tables
\usepackage{color}
\usepackage{array} % for better arrays (eg matrices) in maths
\usepackage{paralist} % very flexible & customisable lists (eg. enumerate/itemize, etc.)
\usepackage{verbatim} % adds environment for commenting out blocks of text & for better verbatim
\usepackage{subfig} % make it possible to include more than one captioned figure/table in a single float
\usepackage{graphicx,url}	% Including figure files
\usepackage{amsmath}	% Advanced maths commands

\usepackage{fancyhdr} % This should be set AFTER setting up the page geometry
\usepackage[nottoc,notlof,notlot]{tocbibind} % Put the bibliography in the ToC
\usepackage[titles,subfigure]{tocloft} % Alter the style of the Table of Contents

\title[Short title, max. 45 characters]{ Hamilton's Object - a clumpy galaxy straddling the gravitational caustic of a galaxy cluster : Constraints on dark matter clumping}

\author[Richard E. Griffiths et al.]{
Richard E. Griffiths,$^{1,2}$\thanks{E-mail: griff2@hawaii.edu (REG)}
Mitchell Rudisel,$^{1}$ 
Jenny Wagner,$^{3}$ \newauthor
Timothy Hamilton,$^{4}$ 
Po-Chieh Huang,$^{1,5}$ 
Carolin Villforth$^{6}$
\\
$^{1}$Department of Physics \& Astronomy, University of Hawaii at Hilo, 200 W. Kawili St., Hilo, HI 96720, USA\\
$^{2}$Department of Physics, Carnegie Mellon University, 5000 Forbes Ave., Pittsburgh, PA 15213, USA\\
$^{3}$Universit\"at Heidelberg, Zentrum f\"ur Astronomie, Astronomisches Rechen-Institut, M\"onchhofstr. 12--14, 69120 Heidelberg, Germany\\
$^{4}$Department of Natural Sciences, Shawnee State University, 940 Second Street, Portsmouth, Ohio 45662, USA\\
$^{5}$Dept. of Physics, Chung Yuan Christian University, 200 Chung Pei Road, Chung Li District, Taoyuan City, Taiwan 32023, R.O.C\\
$^{6}$Department of Physics, University of Bath, Claverton Down, Bath BA2 7AY, UK\\
}

\date{Accepted 2021 May 7. Received 2021 May 7; in original form 2021 Mar 19}

\pubyear{2021}

\begin{document}
\label{firstpage}
\pagerange{\pageref{firstpage}--\pageref{lastpage}}
\maketitle

% Abstract of the paper
%%%
\begin{abstract}
We report the discovery of a `folded' gravitationally lensed image, 'Hamilton's Object', found in a HST
image of the field near the AGN SDSS J223010.47-081017.8 (which has redshift 0.62). The lensed images are sourced by a galaxy at a spectroscopic redshift of 0.8200$\pm0.0005$ and form a fold configuration on a caustic
caused by a foreground galaxy cluster at a photometric redshift of 0.526$\pm0.018$ seen in the corresponding Pan-STARRS PS1 image and marginally detected as a faint ROSAT All-Sky Survey X-ray source. The lensed images exhibit 
properties similar to those of other `folds' where the source galaxy falls very close to or straddles the caustic of a galaxy cluster.
The folded images are stretched in a direction roughly orthogonal to the critical curve, but the configuration is that of a tangential cusp.
Guided by 
morphological features, published simulations and similar `fold' observations in the literature, we identify a third or 'counter'-image,
confirmed by spectroscopy. 
Because the fold-configuration shows highly distinctive surface brightness features, follow-up observations of microlensing or detailed investigations of the individual surface brightness features at higher resolution can further shed light on kpc-scale dark matter properties. 
We determine the local lens properties at the positions of the multiple images according to the observation-based lens reconstruction of Wagner et al. (2019). The analysis is in accordance with a mass density which hardly varies on an arc-second scale (6 kpc) over the areas covered by the multiple images.
% This is a simple template for authors to write new MNRAS papers.
% The abstract should briefly describe the aims, methods, and main results of the paper.
% It should be a single paragraph not more than 250 words (200 words for Letters).
% No references should appear in the abstract.
\end{abstract}

% Select between one and six entries from the list of approved keywords.
% Don't make up new ones.
\begin{keywords}
gravitational lensing: strong;    galaxies: clusters: individual;  (cosmology): dark matter
\end{keywords}

%%%%%%%%%%%%%%%%%%%%%%%%%%%%%%%%%%%%%%%%%%%%%%%%%%

%%%%%%%%%%%%%%%%% BODY OF PAPER %%%%%%%%%%%%%%%%%%

\section{Introduction} 
\label{sec:intro}

Light of all wavelengths responds to the curvature of spacetime, which in turn depends on matter and energy content. One outcome of this response is gravitational lensing, and the study of extragalactic gravitationally-lensed objects has become a very fruitful field of research over the past few decades.  These studies started with the startling discovery of multiply-imaged quasars (Walsh, Carswell and Weymann 1979), and the general subject of extragalactic lensing has since diversified into a number of extremely active and productive areas, including lensing by large scale structures, e.g. the Kilo-degree Survey, KIDS (Troster et al. 2021) and the Dark Energy Survey (To et al. 2020);
 clusters of galaxies (Soucail et al. 1988; the Cluster Lensing and Supernova Survey with Hubble, CLASH (Merten et al. 2015); and the Reionization Lensing Cluster Survey, RELICS (Coe et al. 2019) and individual galaxies: 
 measurement of H$_0$ via time-delay cosmography (e.g. Millon et al. 2020).
Over the last decade, the Hubble Space Telescope has been used to determine the mass distribution in clusters of galaxies via the characterization of the multiple lensed systems found in each of the massive clusters studied (these large projects have included the Hubble Frontier Fields:  Lotz et al. 2017;  the CLASH survey;  and the Beyond Ultra-Deep Frontier Fields and Legacy Observations survey, BUFFALO: Steinhardt et al. 2020).  Lens reconstruction algorithms have been used to infer the masses and other properties of these clusters: see Meneghetti et al. (2017)
 for an overview and a benchmark of common approaches. 
The resulting masses have been compared with the masses inferred from X-ray observations to study the central parts of the cluster and its relaxation state.
Weak gravitational lensing reconstructions in the outer cluster regions complete the reconstruction of the mass density profile of a galaxy cluster.
Constraints on the properties of dark matter (DM) 
afforded by gravitational lensing have been reviewed by Massey, Kitching and Richard (2010).

This paper describes a
strong gravitational lensing configuration, 'Hamilton's Object', consisting of two resolved images of a clumpy spiral background galaxy in a fold configuration and a third resolved image of the same spiral galaxy, which together create a tangential cusp configuration.

 The importance of this discovery is in the potential 
 for constraining small-scale dark matter properties using the highly-resolved multiple image configuration straddling the critical curve. 
Analysis of the individual features observed in the multiple images or the detection of microlensing events on top of the galaxy-cluster-scale magnification effect make such fold configurations ideal probes of dark matter properties. Future goals could include the investigation of the abundance and distribution of possible 
 compact halo objects or ultralight axion dark matter.  

% This is a simple template for authors to write new MNRAS papers.
% See \texttt{mnras\_sample.tex} for a more complex example, and \texttt{mnras\_guide.tex}
% for a full user guide.

% All papers should start with an Introduction section, which sets the work
% in cx
% ontext, cites relevant earlier studies in the field by \citet{Fournier1901},
% and describes the problem the authors aim to solve \citep[e.g.][]{vanDijk1902}.
% Multiple citations can be joined in a simple way like \citet{deLaguarde1903, delaGuarde1904}.

\section{Observations}
\label{sec:obs}

\begin{figure}
	% To include a figure from a file named example.*
	% Allowable file formats are eps or ps if compiling using latex
	% or pdf, png, jpg if compiling using pdflatex
	\includegraphics[width=\columnwidth]{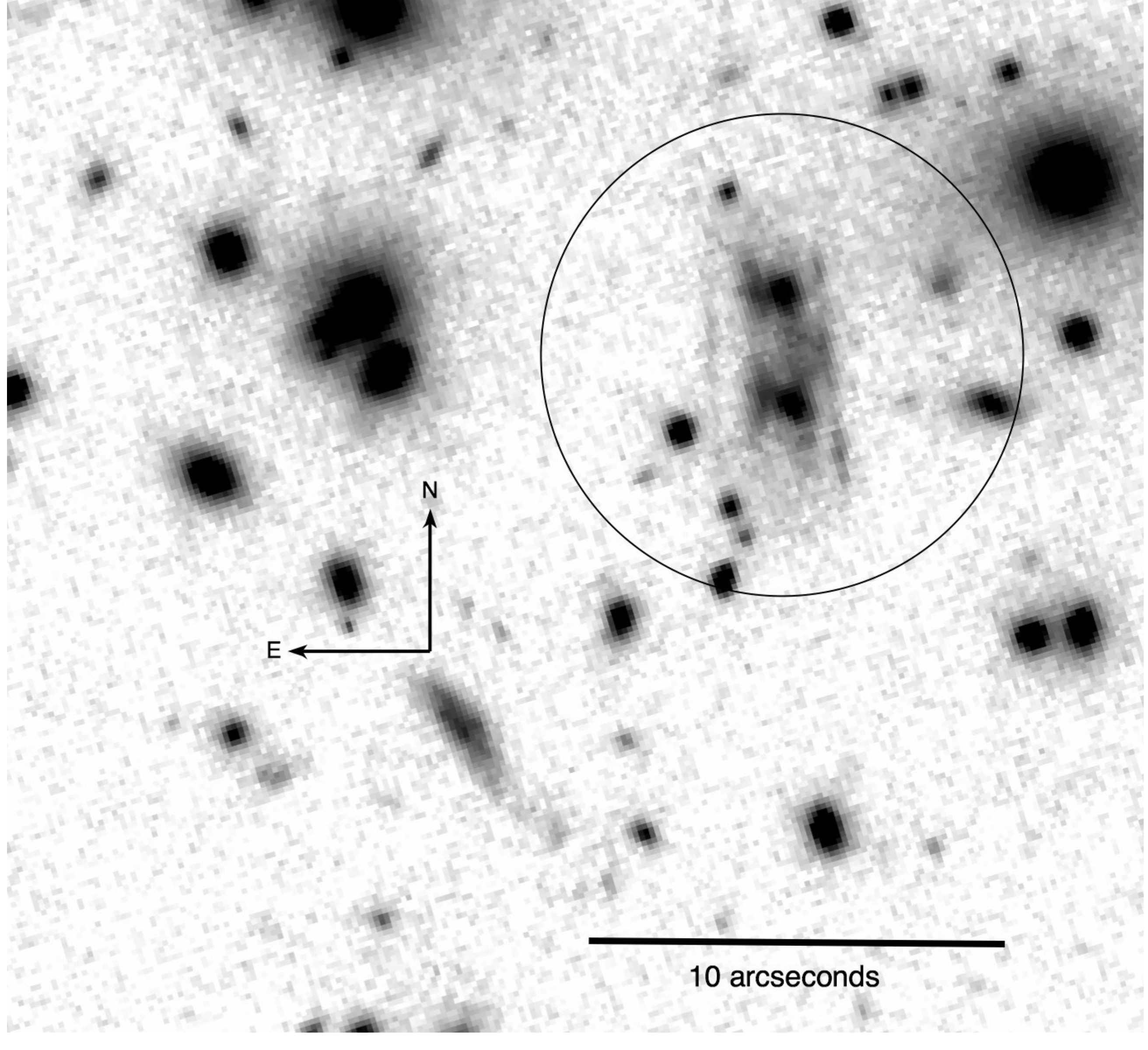}
    \caption{'Hamilton's Object': the HST WFC3  H-band image (3 x 700s). The lensed images are encircled. The halfway point between the two brightest lensed image components is at RA  22h 30m 09.7s, DEC -08$^o$ 09' 40'')
	}
    \label{fig:F110W}
\end{figure}

A complex of folded, distorted  and stretched gravitationally lensed images (Fig.\ref{fig:F110W})  has been discovered in a HST WFC3 image (P.I. Villforth, Program GO 13305) of the field around the AGN SDSS J223010.47-081017.8, an AGN possibly identified with the marginally detected ROSAT All Sky Survey source RASS 1RXS J223011.3-081024 (from Anderson et al. 2007, listed as RASS 6135 in Villforth et al. 2017).
The complex lensed images are surrounded by 
a cluster of galaxies evident in the Pan-STARRS PS1 images archived at the Hubble Legacy Archive (https://hla.stsci.edu), and 
also in the Sloan Digital Sky Survey (SDSS).
The SDSS cluster is catalogued as Red Mapper Cluster RM J223013.1-080853.1 (Ryckoff et al. 2016; http://risa.stanford.edu/redmapper/ ). redMaPPer is an optical cluster finder based on the detection of spatial overdensities of red sequence galaxies (Rykoff et al.
2016, McClintock et al. 2019).  The cluster also appears in the Wavelet Z Photometric (WaZP) catalog (q.v.: Aguena et al. 2021) in which overdensities are found based on redshifts rather than colours.

% x-ray data
% - spectral information (about the three images and about the cluster member galaxies)
% - optical data in various filter bands

\subsection{X-ray Detection of the Galaxy Cluster}
\label{sec:xray}

Thus far, there have been no pointed (or slew-survey) x-ray observations of the cluster using XMM-Newton or the Chandra X-ray Observatory.
But, as well as ROSAT detection of the nearby AGN, the cluster 
was marginally detected in the ROSAT All Sky Survey (RASS) as x-ray source 1RXS ~J222956.9-080823 (Voges et al. 2000) with x-ray countrate
 $ 3.6\pm1.7\times10^{-2} $ counts s$^{-1}$ corresponding to a flux of $3.8\pm1.8\times10^{-13}$ ergs cm$^{-2}$ s$^{-1}$ (0.1 - 2.4 kev).
  The RASS Faint Source Catalogue (RASSFSC) is included in the 
 second ROSAT All-Sky Survey Catalog  (2RXS: Boller et al., 2016 ) but the 2RXS has only one source in the vicinity of the AGN and the cluster, a source which is assigned
  a 43\% probability of extension, with extremely few x-ray counts.
 
%[ The angular distance between the AGN and the centroid of the faint cluster X-ray sourceâ is 4 arc mins. compared with the ROSAT FWHM response of ...  ]
 
 The cluster x-ray source was too faint to be included in the ROSAT-ESO Flux-Limited X-ray Galaxy Cluster Survey (REFLEX 
 Boehringer et al., 2004) or in REFLEX II (Chon \& Boehringer, 2012), even though those papers addressed clusters in the appropriate redshift range.
  % The x-ray counts are insufficient to comment on possible source extension, but the centroid of the counts is within the cluster of galaxies, close to the AGN RASS 6135 at redshift 0.62 (Villforth et al. 2017).
Likewise, the cluster is not included in the Master Catalog of X-ray Selected Clusters, MCXC (Piffaretti et al. 2011).
With a photometric redshift for the cluster of $0.526\pm 0.018$ (redMaPPerClusters - from SDSS DR8:  NED RM~J223013.1-080853.1),  the X-ray luminosity (taking the 1XRS cluster source flux) is $3.0\pm1.5\times10^{44} $ ergs s$^{-1}$ (0.5 - 2.4 keV).  Using the correlation between cluster mass and x-ray luminosity found by Stanek  et al. (2006), this luminosity would imply an approximate mass of 5$\pm3\times10^{14}$ M$_\odot$, in the mid-range for clusters.
   
\subsection{Optical Images}
\label{sec:opt}

\begin{figure}
	% To include a figure from a file named example.
	% Allowable file formats are eps or ps if compiling using latex
	% or pdf, png, jpg if compiling using pdflatex
\includegraphics[width=\columnwidth]{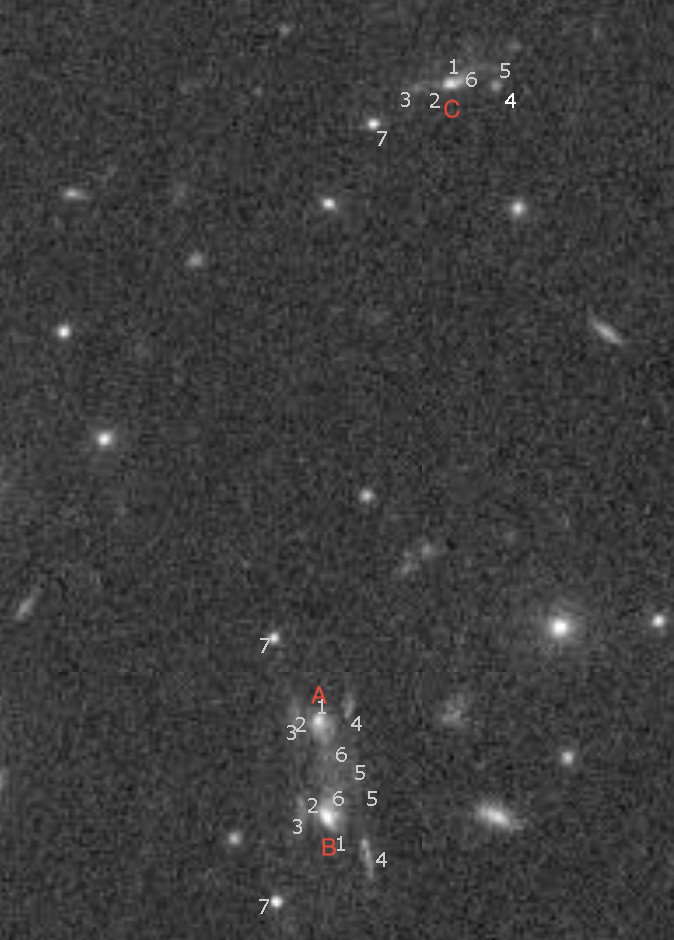}
   \caption{The HST ACS/F606W image (3600 s) showing folded images $A$ and $B$ and third or counterimage $C$
   (see Sec. \ref{sec:candidate}).  
   The nuclear bulges are A1, B1, C1, and the surface brightness features of C are labelled 2 --7, with corresponding
   features in fold images A and B.  These features are seen more clearly in Figs. \ref{fig:C3} and \ref{fig:features}. The overall frame size is $18\times25$ arcsecs.
}
 \label{fig:ACS_ABC}
\end{figure}

The discovery image of 'Hamilton's Object' in the HST WFC3 H-band  is shown in Fig.\ref{fig:F110W}  and the F606W image in Fig.\ref{fig:ACS_ABC}, where the latter image shows 
structure (superstarburst regions) within 0.2 arcsecs. to the west of the nuclear bulge.
The overall object shows a (roughly north-south) mirror symmetry, with two brighter component
images (nuclear bulges), one in each half, and three (split) linear features, spanning a total of about 4 by 6 arcsecs. 
The two brightest central (nuclear bulge) images have F606W magnitudes of $\simeq24$ from SExtractor in the HST archive,
and each has slightly-arced and stretched images to both the east and west. 
Tab.~\ref{tab:mags} 
%(see Appendix)
 shows the Sextractor magnitudes of the central nuclear bulge objects in the HST filters.

There is a very high degree of symmetry about a line running through the center of the whole lensed image at a position angle of about 100 degrees (Figs.\ref{fig:F110W} and \ref{fig:ACS_ABC}).

The original HST WFC3 image (with exposure of $3\times700$s) was made in 2013 with the infrared F160W (H-band) filter
(GO 13305).
Since the original HST WFC3 image (Fig.\ref{fig:F110W}), further observations of this field have been made in 2015 and 2016 under HST 
snapshot programs GO-13671 and  GO-14098 (part of the program of Repp \& Ebeling 2017).
These HST snapshot programs included ACS/WFC observations through filters F606W and  F814W,
 and further WFC3 observations using  filters F110W and F160W. The ACS/WFC exposures were of duration $3\times1200$~s
in each filter and the additional WFC3 exposures in 2016 were $2\times706$~s (F110W) and $4\times1412$~s (F140W). 
The magnitudes of the nuclear bulge components (in Fig.~\ref{fig:F110W} etc.) in HST filters are summarized in Tab.\ref{tab:mags}.
The snapshot observations were targeted on the X-ray cluster `eMACS J2229.9-0808A'
(see https://hla.stsci.edu), corresponding to the marginally-detected  x-ray source 1RXS J222956.9-080823 in the first version of the RASS Faint Source Catalog, RASSFSC.

\begin{table*}
\caption{Nuclear Bulge Magnitudes of Lensed Components (from HLA Sextractor, mag-auto values)}
\label{tab:mags}
\begin{tabular}{cccccc}
\hline
Image   &  F606W &   F814W &   F110W &  F140W  &  F160W   \\
% &   &    \\
\hline
$A$ & 23.17 &  22.46  & 21.55 &  21.29  &  21.10\\
$B$ & 22.80 &  22.20  & 21.22  &   20.93 &  20.86 \\
$C$ & 23.69 &  23.30  &   22.13   &  21.87  & 21.57  \\
\hline
\end{tabular}
\end{table*}

\begin{figure}
\includegraphics[width=\columnwidth]{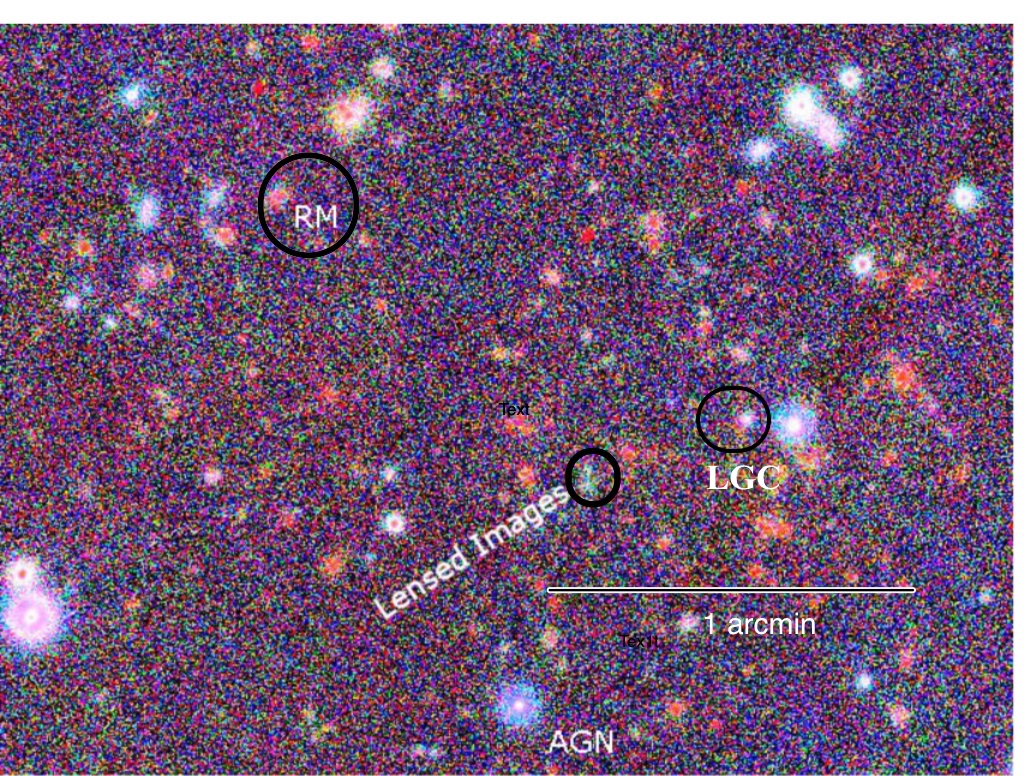}

\caption{Pan-STARRS PS1 combined z-r-g image of the galaxy cluster RM J223013.1-080853.1 (Ryckoff et al. 2016; http://risa.stanford.edu/redmapper/) showing the location of the lensed images.
  RM marks the galaxy found as the cluster 'center' or brightest cluster galaxy by the RM software. The QSO or AGN is also marked, as are the lensed images $A$ \& $B$,
  which are blue and marginally resolved. The approximate location of the local gravitational centre, as determined by lensing analysis (Sec. \ref{sec:lens}) is marked LGC}
\label{fig:PS1}
\end{figure}

\begin{figure}
	\includegraphics[width=\columnwidth]{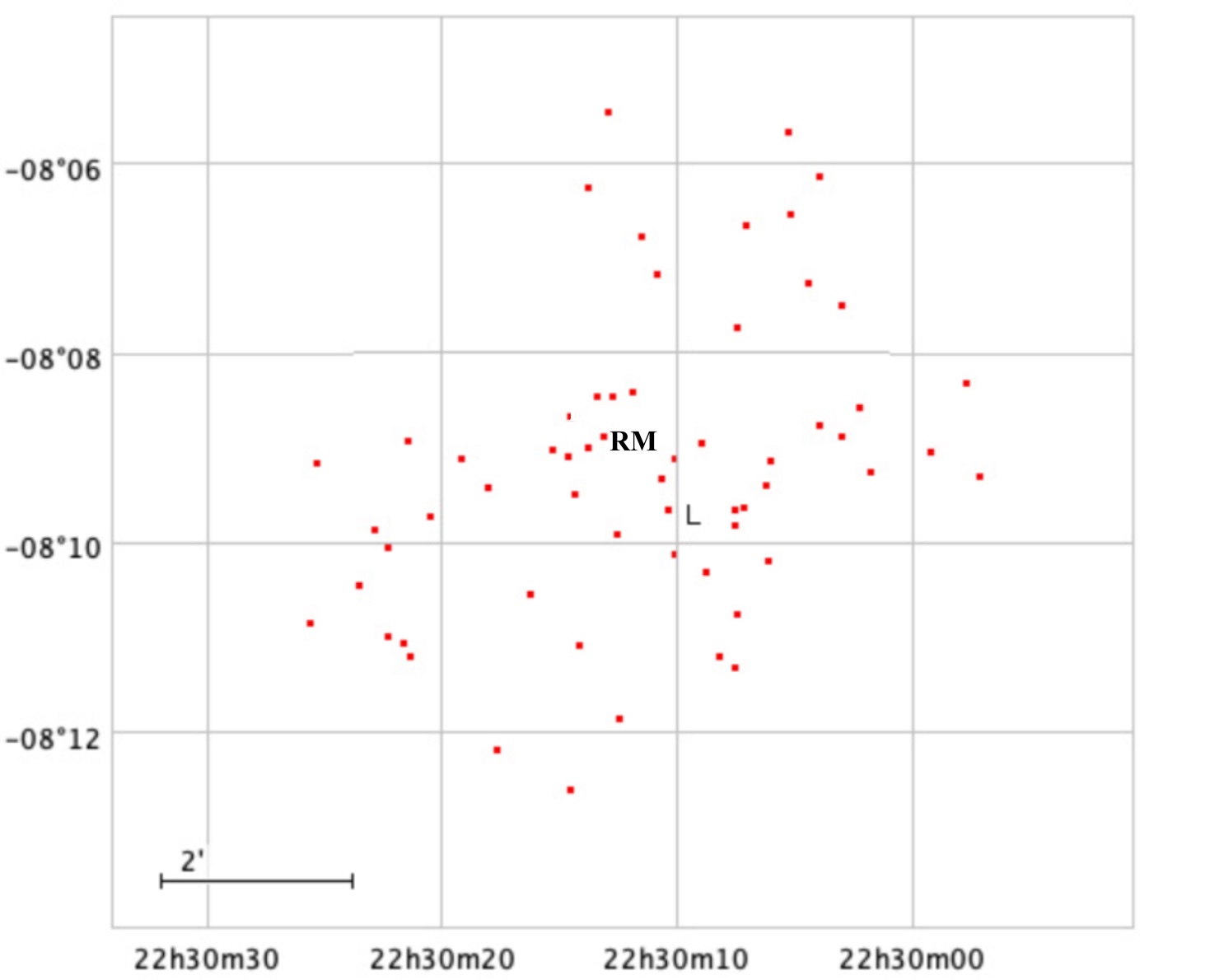}
    \caption{redMaPPer cluster RM223013.1-080853.1 showing locations of photometric cluster members.
    The galaxy labelled RM is the one found by redMaPPer as the brightest.
    The position of lensed fold images $A$ and $B$ (Fig. \ref{fig:F110W}) is marked as L.}
    \label{fig:RM}
\end{figure}

The main nuclear bulge images $A$ and $B$ are resolved in the Pan-STARRS PS1 images in the Hubble Legacy Archive,
and the combined Pan-STARRS z-r-g image is shown in Fig. \ref{fig:PS1}.
'Hamilton's Object' is clearly bluer than the cluster galaxies, i.e. the source % is in the background at higher redshift
 has ongoing star formation, consistent with a spiral star-forming galaxy. 

Surrounding the lensed images is a galaxy cluster for which the apparent members have Sloan Digital Sky Survey (SDSS)
 photometric redshifts averaging $0.526 \pm 0.018$, as measured and catalogued in redMaPPerClusters in the NASA Extragalactic Database, NED:  RM J223013.1-080853.1.
The redMaPPer catalog cluster members  are shown in Fig. \ref{fig:RM} (using the Tool for OPerations on Catalogues And Tables, TopCat (Taylor 2005)). There
is no core- or central-Dominant (cD) galaxy present, unlike the situation in many richer and more relaxed clusters. The catalogued RM cluster centre is positioned on one of the brightest galaxies,
labelled $RM$ in Figs. \ref{fig:PS1} and \ref{fig:RM},  but this 
galaxy is probably not the actual cluster center, as seen from its peripheral position, and also suggested by our multiple image lensing analysis (Sec. \ref{sec:lens}). This problem of identification of the cluster center is in common with many RM clusters (Ryckoff et al. 2016). The total number of $\simeq L^*$ galaxies is about 60;  so, by comparison with other clusters (Andreon and Hurn 2010)
 we might expect the total cluster mass to be weakly constrained to $3 \pm 2 \times 10^{14}$M$_\odot$.
 
\begin{figure}
	\includegraphics[width=\columnwidth]{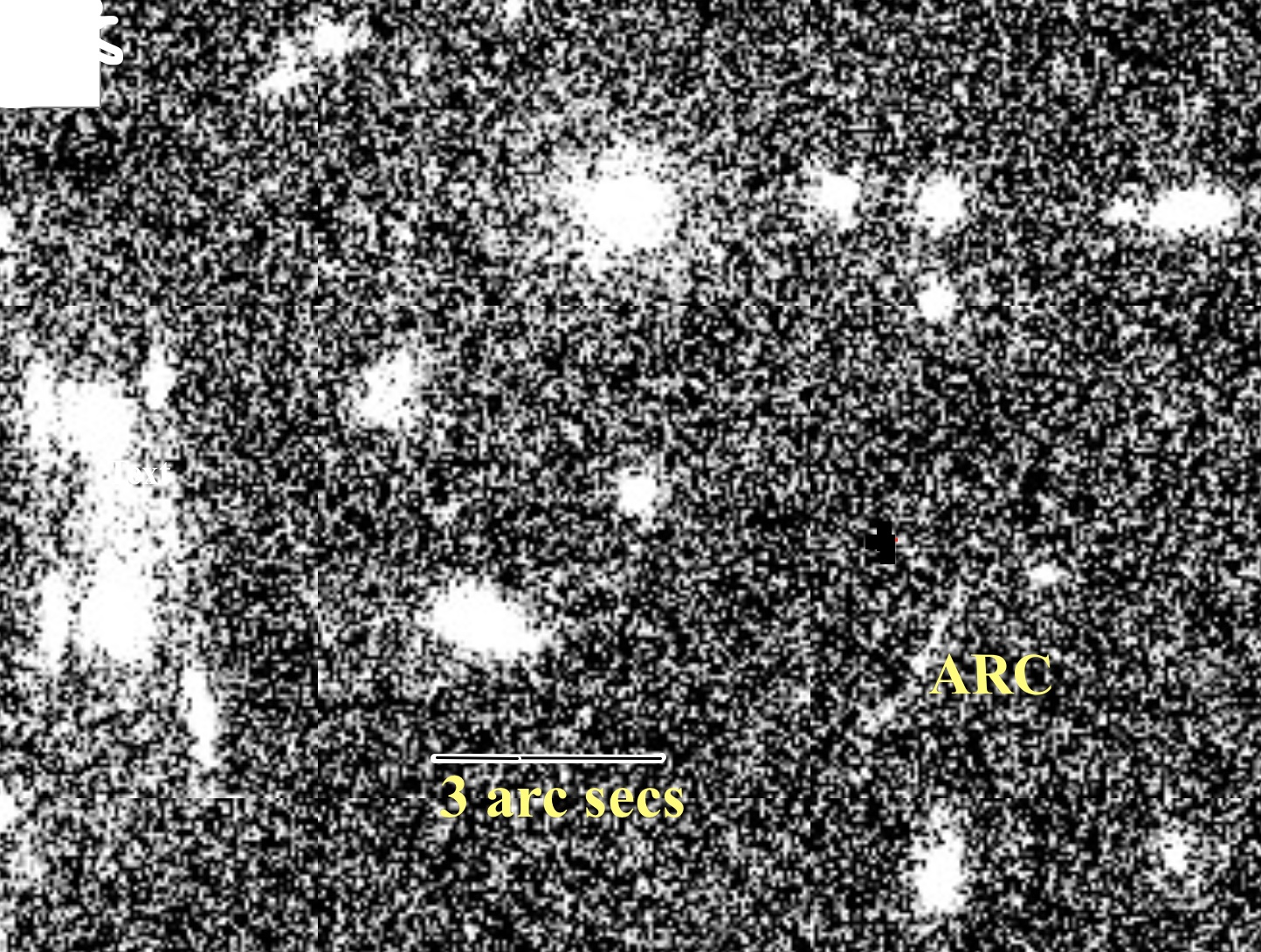}
    \caption{Stretched F606W image showing a faint arc to the west of folded images A/B.}
    \label{fig:arc}
\end{figure}

 A very faint gravitational arc appears in the HST F606 images, centred at RA 22 30 08.97, -08 09 42.0, and even more faintly in F814W (Fig. \ref{fig:arc}).
 The arc is apparently split into three parts and possibly indicates that the center of mass of the cluster is likely somewhere between $A/B$ and counterimage $C$,
 but the faint arc may be additionally lensed by a background galaxy. %it may be on the critical line ...

\section{Finding the Third or 'Counterimage'}
\label{sec:third}

\subsection{Comparison with a Similar Caustic-Straddling Fold in HFF MACS J0416.1-2403 }
\label{sec:MACS}

By comparison with similar objects in the HST Frontier Fields, and in other clusters, and by reference to the theoretical interpretation of Wagner (2019) and Wagner \& Bartelmann (2016), 'Hamilton's Object' is clearly a gravitationally lensed `fold' straddling the critical line in the lens plane
 of a foreground cluster of galaxies (e.g. Keeton, Gaudi \& Petters, 2005, Fig.1; Wagner 2019, Tables 2,3).
 In the absence of a spectroscopic survey of the cluster and background objects, or spectroscopy of the lensed images $A$ and $B$ (Figs. \ref{fig:F110W} and \ref{fig:ACS_ABC}), the HST field images were examined to identify possible candidates for the third or 'counterimage' or other lensed image components. 
 
\begin{figure}
\includegraphics[width=\columnwidth]{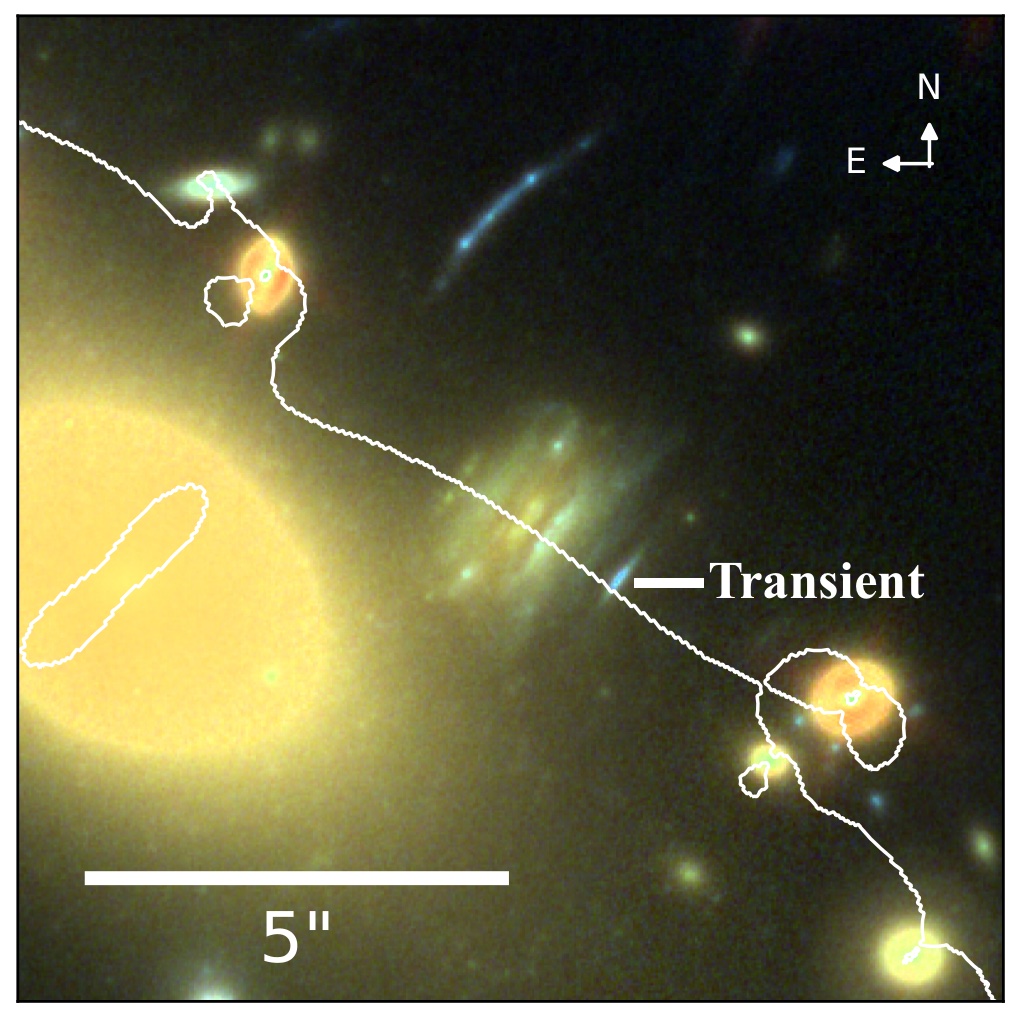}
 \caption{Reproduction of Kaurov et al.(2019) Fig.1: Combined false-color image using the F105W,
F814W and F606W filters showing the region
centered on lensed system \#12(a,b) in Caminha et al. (2017) in MACS
J0416.1-2403. The white line is the calculated critical curve for a source redshift z = 0.94 (Caminha et al. 2017) which almost coincides with the isodensity contour for a projected mass column density of $2 \times 10^{15}$ M$_\odot$ Mpc$^{-2}$. The transient source (Kaurov et al. (2019), Chen et al. (2019)) appeared within the blue arc indicated.  }
\label{fig:Kaurov}
\end{figure}

 The lensed objects in Fig.\ref{fig:F110W} can perhaps best be understood by comparison with lensed system \#12 in the Caminha et al. (2017) paper on the Hubble Frontier Field (HFF) cluster MACS J0416.1-2403. Caminha et al.'s system \#12 is also system \#28 in Jauzac et al. (2014) and system \#35 in Diego et al. (2015).
In Caminha et al.'s  system \#12, the source galaxy (an irregular, 'stringy' or 'clumpy' galaxy) has a redshift of 0.94 and its images straddle the corresponding critical line that envelopes the central part of the cluster MACS J0416.3-2403 at redshift z = 0.396.  Remarkably, an observation has been made that a supergiant on the outskirts of the galaxy has been microlensed and has clearly been greatly magnified during the two years covered by HST observations as it crossed the caustic (discovered independently by Chen et al. (2019) and Kaurov et al., (2019)); the peak magnification factor was about 1000 when all HST observations are included (Chen et al. 2019). 
This transient is within a small fraction of an arcsecond from the critical line (Fig.\ref{fig:Kaurov}).
In the same HFF field, Rodney et al. (2018) detected a pair of transients at two separate locations in a highly magnified galaxy at z = 1.01.

In an earlier example of this remarkable microlensing phenomenon, Kelly et al. (2018) discovered a transient lasting several weeks in a highly magnified galaxy at z = 1.49 in another HFF field, MACS J1149.5+2233.The transient object (?MACS J1149 Lensed Star 1 (LS1)?) was magnified by more than a factor 2000.
 Venumadhav, Dai  \& Miralda-Escude (2017) have shown the 
importance of such microlensing observations for investigations into the substructure of dark matter.
A mass fraction in microlenses disrupts the otherwise smooth caustic into a network of corrugated microcaustics and the light curve of a caustic-crossing star 
is thus broken up into numerous peaks.  The sensitivity of caustic-crossing events to
the granularity of the lens-mass distribution makes them ideal probes of dark matter structural components,
such as compact halo objects or ultralight axion dark matter. 
In a similar context, Dai et al. (2020) have commented on the possibility that asymmetric surface brightness structure of the caustic
crossing arc SGAS J122651.3+215220 in the cluster SDSS J1226+2152 may be evidence for dark matter substructure with subhaloes of
$\sim10^6 - 10^8$~M$_{\odot}$.
  Further, Dai (2021) has shown that star clusters, microlensed in caustic crossings,
can result in observable variability in HST and JWST data and that such data can potentially be used to constrain the substructure of the DM. 
Any flux anomaly in strongly lensed systems can be used to probe subhalos
in the lensing galaxies (Dalal \& Kochanek 2002), and the promise of this technique in future observations from space has been investigated further by Diego (2019).
Dai and Miralda-Escude (2020) have shown that axion minihalos in galaxy clusters should produce subtle surface density fluctuations of amplitude
$\sim 10^{-4}$ to $10^{-3}$ on projected length scales of $\sim10$ to $10^4$ AU, which imprint irregularities on the microlensing lightcurves of
caustic transiting stars. As summarized in the 'fuzzy' or 'wave' dark matter review by Hui (2021), this effect was also used by Gilman et al. (2019)
to constrain warm dark matter and by Schutz (2020) to limit fuzzy dark matter.
% The interest in these observations stems from the fact that fuzzy dark matter, composed of a non-relativistic Bose-Einstein condensate (Hui et al. 2017), may solve the problem of the missing dwarf galaxies, as shown in the simulations by Schive, Chiueh and Broadhurst (2014)}.

 %%% added
Unlike Caminha et al.'s system \#12, however,
the bulk of the source galaxy image projected as Hamilton's Object is not bisected by the corresponding critical curve. The critical curve runs alongside the outskirts of the galaxy,
and through an outer spiral arm -- see Sec.~\ref{sec:analysis}. 
% (see Sec.\sec{lens} and Fig.\fig{Hera}. Nevertheless, stars will certainly cross the critical curve.

% These events
% were identified during two month-long campaigns to image MACS0416 as part of HFF project (PI J. Lotz).
% While the events each lasted only several
% weeks, their interpretation was not immediately apparent.
% The detection of the lensed star in MACS J1149 magnified by > 2000 at peak brightness prompted the
% interpretation of the two MACS0416 events as likely microlensing events (Rodney et al. 2018).

%Other such transient behaviour has been seen in other lensed objects (e.g. Rodney et al. 2018).

The HST observations of this field were taken over a time span of 3 years (2013 - 2016). There were no discernible changes to the images, in contrast with system \#12a in Caminha et al.
 
 Nevertheless, Hamilton's Object may be one of the few additional candidates, thus far, for the potential future observation of microlensing of individual stars on the critical line of a cluster of galaxies. 
 
In  Caminha et al.'s system \#12, the critical line for z = 0.94 bisects the folded galaxy image (Fig. 1 of Kaurov, reproduced here as Fig. \ref{fig:Kaurov} ) and almost happens to coincide with the isodensity contour for $2 \times 10^{15} M_{\odot} Mpc^{-2}$, shown as the white line in Fig. \ref{fig:Kaurov}. 
There should then be a third, sometimes called a 'counterimage', which is identified as object \#12a of Caminha et al. Unfortunately, this image is on the northern side 
of an unassociated galaxy (Caminha et al. Fig. A1) and may be  partly confused with the outer spiral arm of that galaxy, but the MUSE imaging/spectroscopy
shows that it is a separate object. This third image shows that the source object is a clumpy irregular galaxy.
In Diego et al. (2015), the authors did not have access or use the 
archived MUSE data of Caminha et al., so Diego et al. (2015) made two incorrect guesses on the counterimages, each of them  at the wrong redshift, so their attempted modelling for this lens is a poor fit to the data.  Due to the very elongated
nature of the cluster mass distribution in MACS J0416.1-2403 (Richard et al 2014, their Fig. 4), many or all lensed systems 
appear as triply-imaged configurations (e.g. systems \#4,~ \#5,~\#7,~\#13,~\#16 in Caminha et al. 2017) , but  system ~\#12 summarised above may be one of the few systems on the critical line for its redshift. All of the lensed systems have been used jointly to determine the mass distribution in MACS J0416.1-2404 (Zitrin et al. 2013; Grillo et al 2015, Jauzac et al. 2014, Caminha et al. 2017, Diego et al. 2015) using the LensTool software developed by Julio, Kneib et al. (2007) or other software packages.

Other examples of gravitational folds and cusps include the VLT/MUSE Deeply Lensed Field where a Lyman-$\alpha$ emitter at z = 6.6
 has been found by Vanzella et al. (2020) and is also on a critical line for its redshift.  Further examples of triple systems can be found in RELICS ( Cerny et al. 2018, Coe et al. 2019). A more extreme example, with a background galaxy lensed into an Einstein ring by a cluster, is that of Abell 3827 (Chen et al. 2021) where the authors show that dark matter is required over any alternative gravitational conjecture.

\subsection{The Candidate Counterimage}
\label{sec:candidate}

By comparison with the above observations of lensed systems in MACS J0416.1-2403  and other observations 
of similar systems (e.g. the folded lensed object in galaxy cluster MACS J1149.5+2223;  
Wagner \& Bartelmann 2016 (WB16), their fig.4), together with simulations (Wagner 2017) %A\&A astro-ph 1612.01793)
  there should be a third or 'counterimage' thrown off roughly in the direction of the elongation of the stretched clumps and towards the lens center, where
  the latter has not been determined by the photometric surveys or any other method.
The simulation by Wagner (2017, fig.4), however, shows that the lens center is on the opposite side of the critical line 
to the more `splayed out' or distorted part of the folded images, despite the simplicity of the simulation.
% This simulation shows a simple, circular lens that is not able to produce the observed confiiguration: the NFW profile can only produce radial arcs at the innermost radial critical curve; this lens model is usually too simple for realistic clusters such as the subject of this paper.
 We note that the southern part of Hamilton's Object
is the more splayed;  so the lens center, as well as the third or counterimage, can be expected to be found on the northern side of the two folded  images.
 Further, by comparison with the lensed system in MACS J1149.5+2223 (Wagner \& Bartelmann 2016, fig.4)
 and also with system \#12a of Caminha et al. (2017), we might expect to find the third or counterimage
 of Hamilton's Object at a distance from the critical line of about five to ten times the angular distance 
 between the two nuclear bulge components of the folded pair, which are separated by 2.6 arcsecs.
 (e.g. Narayan \& Bartelmann 1996, Wagner \& Bartelmann 2016).
 
Based on position, colours and morphology, we thus identified a candidate third or 'counterimage' as $C$ ~in Fig. \ref{fig:ACS_ABC},
 a `clumpy' disk galaxy seen nearly edge-on.
  We know rather little  about the cluster morphology, 
however (see Figs. 3, 4), so it might be that the cluster is an elongated merger similar to those mentioned above. A Singular Isothermal Ellipsoid (SIE) would lead to a fourth image on the other side of the cluster center, and we have an unconfirmed candidate for that, but further observations are required (we assume here that the galaxy marked $RM$ in Fig.\ref{fig:RM} is not at the cluster centre). 

The nucleus and clumpy galaxy are lensed into a `fold' configuration straddling the critical line in the lens plane, and the components are therefore stretched in a direction roughly orthogonal to the critical line (e.g. WB16), in the same way as objects \#12 b, c of Caminha et al. in EMACS J0416.1-2403, or the folds in the cluster SDSS J1226+2152 (Dai et al. 2020).
 The stretched arcs are formed as two images of the source galaxy merge at the position of the critical curve, which results in
the mirror-symmetric appearance. The magnified arcs constrain the local position and shape of the critical curve, as shown below.

The relaxation state of the galaxy cluster RM223013.1-080853.1 has not been determined yet, and Hamilton's Object may be the result of an ellipsoidal mass density profile or of a merger structure like EMACSJ046.1. In both configurations, a third counterimage is expected north of Hamilton's Object as detailed previously. The direction of `stretch' or elongation in the outer (East and West) lensed components is not orthogonal to the apparent line of symmetry or critical line (Figs. \ref{fig:F110W} and \ref{fig:ACS_ABC}).
In fact, the direction of elongation is at an angle of about 20 degs. to the orthogonal to the line of symmetry or critical line. This rules out a simple Navarro-Frenk-White (NFW, Navarro, Frenk and White 1997) radial mass profile for the cluster (Wagner 2019b), in favor of an ellipsoidal profile
more consistent with the observed (irregular) galaxy distribution (Fig. \ref{fig:RM}). In section 4 below, we show via detailed simulations that the NFW profile is ruled out.
In the massive cluster MACSJ0416.1-2403, the isodensity contours are very elongated, because the cluster is the merger of two large sub-clusters, and system \#12 of Caminha et al. and other systems lensed there are close to the very elongated critical line (see Fig. 4 of Caminha et al).
%*** SEE NOTE BELOW 

\begin{figure*} % [h!]
\centering
\includegraphics[width=\textwidth]{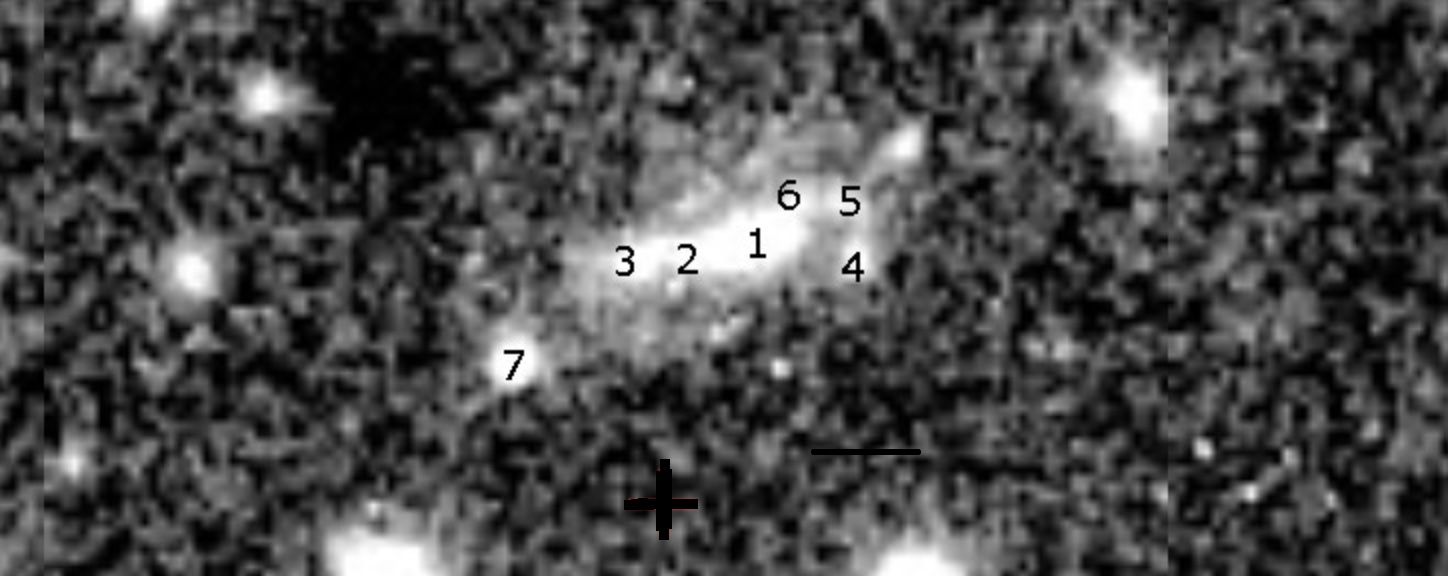}
\caption{The contrast-stretched F110W HST image of 'counterimage' $C$, showing feature $7$ on the edge of the galaxy, which is approx. 4 arcsecs across.
Feature $1$ is the nuclear bulge, and features $2$ through $6$ are also shown in Figs. \ref{fig:features} and \ref{fig:cc}. The contrast has been stretched such that 
 noise is visible at low levels.}
\label{fig:C3}
\end{figure*}

Prior to the spectroscopic observations described below, we had thus identified a candidate counterpart for the third or counter-image as a large disk-like `clumpy' galaxy 
at  RA 22h 30m 09s.45,  dec. -08d 09m 20s.6  (Fig. 2, object $C$), with a nuclear bulge of magnitude 23.7 (in HST ACS/F606W, possibly confused with circumnuclear emission).  Object $C$ is fainter relative to objects $A$ and $B$ as expected ($A$ has nuclear bulge mag. 23.2 in F606W, and $B$ has mag. 22.8). 
This candidature was based on the colour, morphology and astrometry
of the counterimage candidate $C$ components in both visible and the infrared bands: the characteristics were matched with components of the lensed images $A$ and $B$ in 
both bands. The nuclear bulge and `clumps' in image $C$  are stretched radially in lensed objects $A$ and $B$. In addition, $C$ has an arc-like feature to the NW of the nuclear bulge and this appears, folded, between
 and to the west of bulge images $A$ and $B$, and the two 'mirrored' faint arcs meet close to the critical line. 
 The faint arc-like feature is interpreted as a spiral arm of the disk seen at low inclination (Fig. \ref{fig:C3}). In image $B$ the parity is  
'positive' (even); in image $A$ the parity is negative or odd, i.e. the counterimage $C$ is flipped around an E-W axis roughly parallel to the critical line.

 In the HST F606W images, the morphology and astrometry of the many components in
  candidate $C$ clearly match the 
 folded and stretched morphologies of the components in $A$ and $B$. 
 It is not immediately obvious that the compact object
 labelled C7 in fig. \ref{fig:ACS_ABC} is physically associated with galaxy $C$, but a stretched-contrast image of the galaxy in F110W (Fig. \ref{fig:C3}) makes this association extremely likely, and the object may be an outlying superstarcluster. In this case, the counterimage outlying object C7 appears as objects A7 and B7 in the folded images. 
 
% In this context, we note that object B2 has brighter areas on the north side, in contrast with the southern
% edge of object A2, in further support of the conclusion of Dai et al. that such differences are caused by dark matter substructure.  We defer a detailed analysis of the asymmetric surface brightness structure to a later publication. 
 
 % Another, less likely possibility, is located at  RA 22 30 09.68,   dec -08 09 16 in the F606 image (Fig.     b.).
With regard to the nature of the source galaxy, we note that there are no shortages of `clumpy' or `chain'  galaxies at moderate to high redshifts, as shown from many HST surveys (e.g. HST Medium Deep Survey: Naim et al. (1997), 
%Conselice et al. (2003),
 Abraham et al. (1996); `Chain Galaxies' (Cowie et al. 1995),  Buta (2011); the HDF: van den Bergh, et al., 2000, Elmegreen, Elmegreen and Sheets (2004); GOODS: Elmegreen and Elmegreen (2006), 
 Elmegreen et al. (2009); COSMOS: Scarlata et al. (2007); CANDELS:  Peth et al. (2015); Lotz, Primack and Madau (2004). 
 It has been shown that `chain' or stringy galaxies are usually clumpy disk-like galaxies seen at low inclination angles (e.g. Dalcanton and Shectman 1996; 
 see also Elmegreen et al. 2021)
 and this is apparently the case here.  By stretching the contrast of the HST archival images (Fig.\ref{fig:C3}), we note that
 $C$ is a clumpy disk galaxy seen at low inclination, with a bright nuclear bulge. The overall angular size of image $C$ is at least 4 arcsecs, corresponding to 30 kpc at a redshift of 0.82 (from Ned Wright: Cosmology Calculator I, $H_0=69.6$, $\Omega_M = 0.286$, flat universe ). This is, of course, the apparent size after magnification
 and distortion by the cluster lens.
 
 Having established the clear candidature of galaxy $C$ as the counterimage, spectroscopy was performed as described in section 3.3 below, 
 confirming it. The counterimage $C$ is the image best representing the morphology of the background source because it is apparently the least distorted image.
 
% {\bf Lens Modelling using `lenstronomy'} [to be completed]

% Using the counterpart identified as image `C' in Fig. 6  (RA 22h 30m 09.45s, dec -08d 09m 20.6s) we have used the `lenstronomy' software package
% (Birrer \& Amara 2018), to try to make a match with the observed gravitationally folded images `A' and `B'
% (nuclear image `A' has coordinates RA 22h 30m 09.73s, -08d 09m 38.9s; nuclear object `B' has coordinates RA 22h 30m 09.69s, dec. -08d 09m 41.5s) .
% The source, with morphology of image `C', is assumed to be located just inside the astroid caustic of the Singular Isothermal Ellipsoid (SIE) cluster, fairly close to a cusp. see , e.g., Narayan \& Bartelmann 1995.
% The cluster of galaxies is assumed to have a mass of about $ 5 \times 10^{14}  M_\odot $, and an ellipticity of 
% 0.3 (??) - see Fig. 3. Taking source image `C' at redshift 0.65 and running the light through the mass profile of the cluster
% at redshift 0.53 
% produces an image as in Fig.????.  We can also take the observed images `A', `B', and run them backwards through
% the cluster to form Fig. ???? which can be compared with image `C'. 

\subsection{Optical Spectroscopy}
\label{sec:spec}
 
\begin{figure}
	\includegraphics[width=\columnwidth]{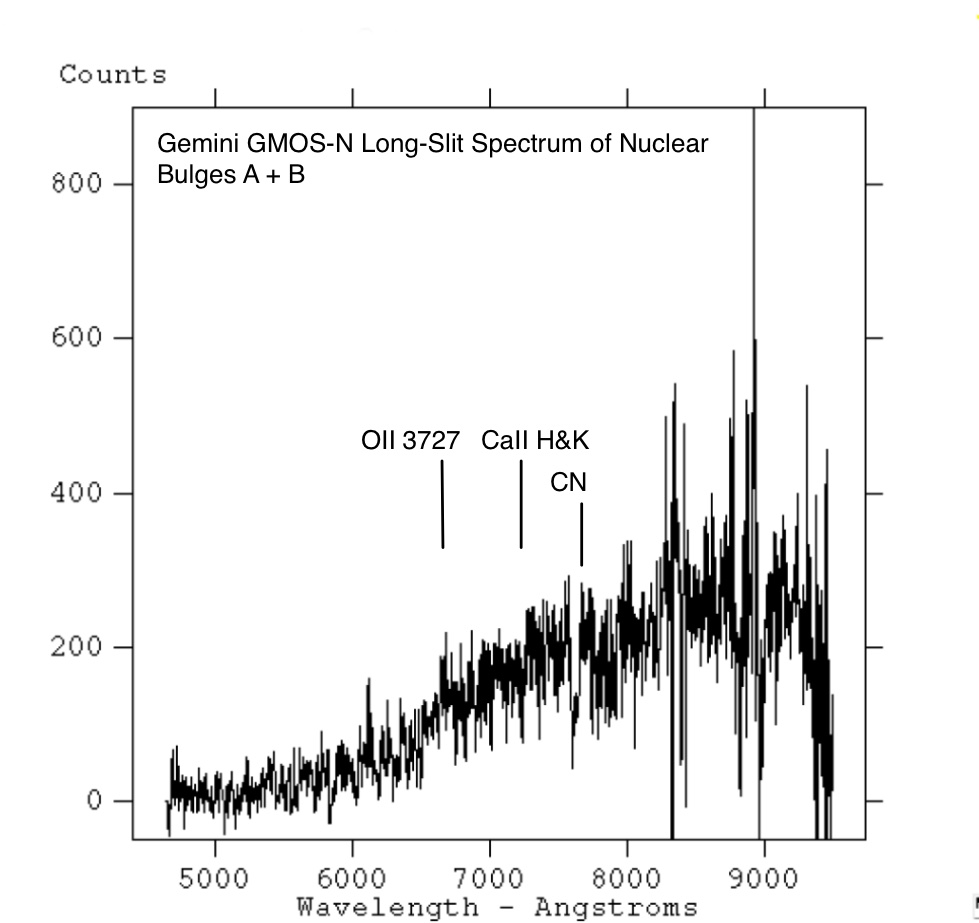}
    \caption{Gemini GMOS-N long-slit spectrum of nuclear bulge images $A1$ and $B1$.   
	}
    \label{fig:GMOS}
\end{figure}

 The lensed and folded nuclear bulge objects $A1$ and $B1$ were observed on Sept 7, 2018, using 
 the Gemini Multi-Object Spectrograph GMOS-N (an integral field unit, IFU) in long-slit
mode (for a description of GMOS-N, see Allington-Smith et al. (2002) and Hook et al. (2004)). The observations used the GG455 filter and the R150 grating disperser centred at 680nm with the red-sensitive, fully-depleted 
Hamamatsu CCDs (the same as those installed in GMOS-S - see Gimeno et al. (2016)), having a plate scale of 0.0807 arcsecs per pixel. Exposure times were of duration 900 secs and the total integration time was 1.5 h.
The resulting summed spectrum of nuclear bulge images $A1$ and $B1$ is shown in Fig. \ref{fig:GMOS}; although the spectrum is of rather low signal-to-noise, 
some features are clearly visible, as summarised in 
%Appendix 
Tab.\ref{tab:GMOS_lines} 
We identify OII $\lambda$~3727\AA, the CaII H\&K break at $\lambda$  4000~\AA~  and CN absorption at  $\lambda \simeq$ 4180 \AA ,
 giving a tentative redshift of 0.82. The long-slit observation
did not include counterimage $C$, and prior observations in 2017 using GMOS-N in full IFU mode did not result in good data because of poor observing conditions. 

\begin{table*}
\caption{Spectral Lines from Gemini GMOS-N data: Objects $A1$, $B1$ only}
\label{tab:GMOS_lines}
\begin{tabular}{lccr}  \\
% &   &    \\
\hline
Ion  & OII  &  CaII H\&K   & CN  \\
Rest $\lambda$ \AA &  3727 & 4000  & 4180 \\ 
Observed $\lambda$ \AA  & 6783 &  7280  & 7608    \\
\hline
\end{tabular}
\end{table*}

\begin{figure}
	\includegraphics[width=\columnwidth]{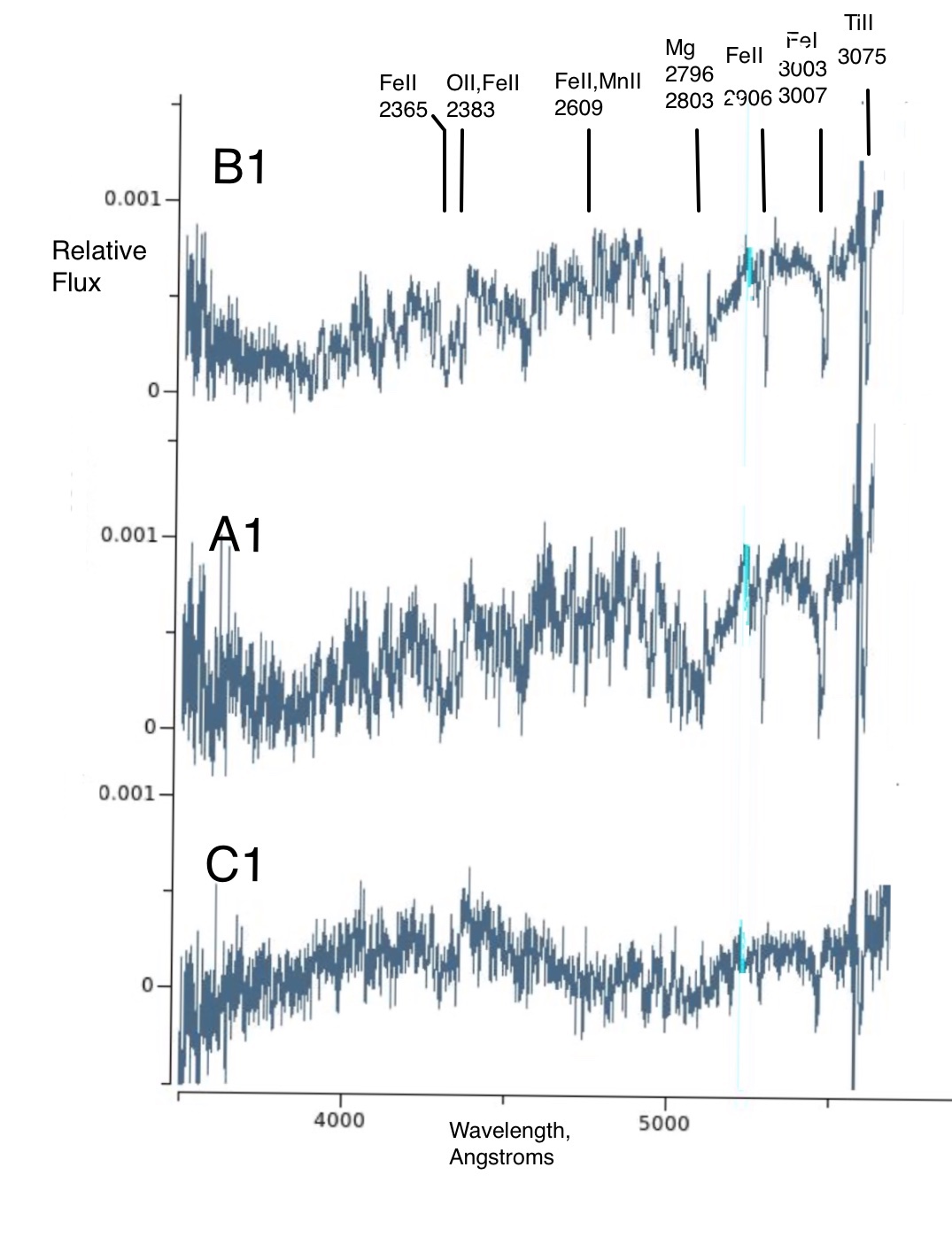}
    \caption{KCWI spectra of the nuclear bulge images 
   $A1$, $B1$ and $C1$ in Fig. \ref{fig:ACS_ABC}. The most prominent lines are labelled with their rest wavelengths - see Table
   \ref{tab:KCWI_lines} for all identified lines.  The line at 5588~\AA~ is a night-sky line.
	}
    \label{fig:KCWI}
\end{figure}

The Keck Cosmic Web Imager, KCWI (see Morrissey et al. 2018) is an IFU that was used at the right Nasmyth focus of the Keck II telescope to make observations of the lensed images on Sept. 16, 2020 UT. The field included all identified 
lensed components $A$, $B$ and $C$. The Blue KCWI was used with the 'large' $8 \times  22$  arcsecs. field of view, with the grating centred at 4500~\AA~ and no red (blocking) filter.
The long dimension of the IFU was oriented N-S so that objects $A$, $B$ and $C$ were covered by a single field, which was observed over 3 hours
with individual exposures of 900s each, in atmospheric seeing that was typically 0.3 arcsecs. The effective spectral coverage extended from 3600~\AA~ to 5600~\AA~ - this was not optimal for an object at a suspected moderate redshift, but was used for schedule and programmatic reasons.
Spectra of the main (nuclear bulge) components  $A1$, $B1$, and  $C1$
are shown in Fig. \ref{fig:KCWI}, and the absorption line identifications are listed in Tab.\ref{tab:KCWI_lines}. The most useful spectral features in the rest-frame NUV
are the resonant absorption lines from the low-ionization species Mg II,  Fe II, CII and OII in
the interstellar medium (ISM) and/or circumgalactic medium(CGM) (Zhu et al.2015). The compilation of absorption line indices in the UV by Maraston et al. (2009) is also
very useful in the identification of NUV absorption lines. 
 All three nuclear bulge images $A1$, $B1$, $C1$ show four strong absorption lines longward of 5000\AA, confirming  galaxy $C$ as the 'counterimage' to the folded 
objects $A$ and $B$. The MgI doublet 2800\AA~ absorption shows a blue wing, perhaps indicative of ouflowing gas 
from the star-forming nuclear bulge, 
in common with this same feature found in many or all large star-forming spiral galaxies at slightly higher redshift, with outflows of typically
 several hundred and up to 1000
km s$^{-1}$  (Wiener et al. 2009). This broad absorption feature may also include discrete absorption lines of MgI. Shortward of 5000\AA,
additional absorption lines observed in $A1$, $B1$ and $C1$ include several lines from FeII, CII  etc. (Tab.\ref{tab:KCWI_lines}),
 but the spectrum of nuclear bulge $C1$ is noisy
in this region. For the gravitationally magnified images $A1$ and $B1$, a smaller  aperture was used to isolate the nuclear bulge spectrum using SAO-IMAGE DS9; this is more difficult
for the counterimage nuclear bulge $C1$, where a larger aperture is needed for better signal-to-noise, so the continuum is not a good match to those of $A1$ and $B1$ at all wavelengths.
The spectra also show that the counterimage galaxy, a large spiral with a compact nuclear bulge, has absorption lines found in some Seyfert galaxies
and LINERS (Low Ionization Nuclear Emission Line galaxies), as noted in Tab.\ref{tab:KCWI_lines}, along with outflowing gas.   

\begin{table*}{l}
\caption{Spectral Lines from KCWI data: Nuclear Bulge Images $A1$, $B1$, $C1$ }
\label{tab:KCWI_lines}
\begin{tabular}{lcccccccccr}
\hline
  Objects &   &   &   &   &    &  & &&\\
% &   &    \\
\hline
Ion                                       &  FeII  &  FeII  &  FeII*  &   CII  &   FeII  & FeII   & FeII, OII  &         &          & FeII, MnII    \\   
 Rest $\lambda$ \AA           & 2249 & 2261 &  2280 & 2326 & 2343   &  2365 &  2383     & 2451& 2459 &    2609    \\ 
  Observed $\lambda$ \AA  & 4094 & 4115 &  4150 & 4230 & 4265 & 4305  &  4350       & 4460 & 4476 &   4747\\
\hline
\end{tabular}
\vskip 0.07in

\begin{tabular}{lcccccccccr}
\hline
 Objects &    &   &   &   &  && &   &&  \\
\hline
Ion                                      &             &          &           &          &    MgI (doublet)   &       MgI  &   FeII      &   FeI (doublet)  &   FeI      &  Ti II, note (a)  \\
Rest $\lambda$ \AA.          &    2676 & 2712 & 2753 &  2766 &                  2796  &      2803 &  2905       &              3004  & 3007     & 3074    \\ 
Observed $\lambda$ \AA   &   4870 & 4935  &  5010 & 5035 &                 5089   &      5101 &  5287      &             5465     & 5473     &  5600    \\
\hline
\end{tabular}
\vskip 0.07in

\raggedright Notes: 
% a) as in N3227 (ref. XXX) ; 
a) as in MCG 8-11-11 (Ulrich et al. 1988)
\end{table*}

The KCWI spectra show that all three nuclear bulge images $A1$, $B1$ and $C1$ have a redshift of 0.8200$ \pm 0.0005$, confirming the Gemini GMOS-N spectrum above. 
The spectrum of counterimage object $C7$,  lensed as images $A7$ and $B7$
also displays the same absorption lines at a redshift of 0.8200.
Given the appearance of $C7$ on the edge of counterimage $C$ (Fig. \ref{fig:C3}), we conclude that this is a superstarburst region or a starburst clump in an outer arm of the galaxy,
similar to an object in Abell S1063 at redshift 0.6 (Walth et al. 2019) or the starburst clump in NGC2207 (Kaufman et al. 2020) . Such ~kpc-sized starburst HII clumps are 
high redshift counterparts 
to  actively star-forming clumps in nearby massive turbulent disk galaxies (Fisher et al. 2016, Elmegreen et al. 2021).  
 
\section{Analysis of the image configuration and morphology}
\label{sec:analysis}

\begin{figure*} % [h!]
\centering
\includegraphics[width=0.6\textwidth]{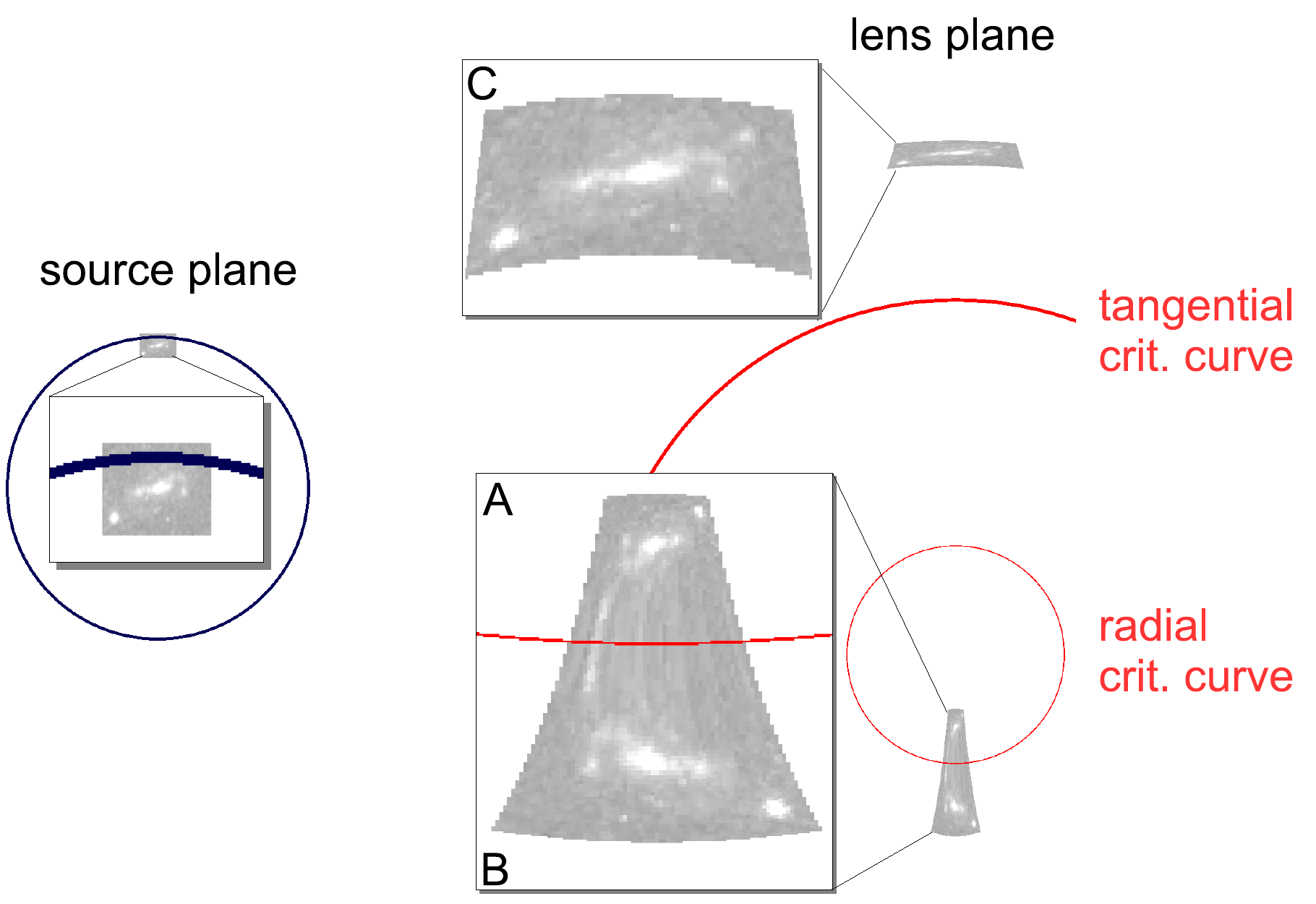}
\caption{Simulation of the multiple-image generation using an NFW profile: mimicking the source with the observed image~$C$ and placing it close to the caustic (blue circle) of an NFW profile (left), this NFW profile gravitational lens generates a triple-image configuration consisting of a counterimage~$C$ outside the tangential critical curve (red arc) and a fold configuration with images~$A$ and $B$ straddling the radial critical curve (red circle) (right). But the parities of the observed images do not match the simulation (see text) - the NFW model is too simple and does not match the observations}
\label{fig:NFW}
\end{figure*}

To further investigate the global cluster shape, we consider mass density profiles which may be capable of generating three images of the same background source showing a high degree of alignment like our observed triple-image-configuration.
An axisymmetric NFW mass density profile is one possible choice. 
This is capable of generating three images of the same background source in the observed configuration, when the source is located inside the caustic curve and 
asymptotically approaching it. 
Focussing on the morphology of the generated multiple images, attention is not paid here to the creation of the exact lens and source positions along the line of sight when simulating an NFW profile. 
Fig.~\ref{fig:NFW} (left) shows its caustic.
Using image~$C$ of the observed multiple image configuration to play the role of the background source at the position shown in Fig.~\ref{fig:NFW} (left), this NFW-profile gravitational lens generates the triple-image configuration as shown in Fig.~\ref{fig:NFW} (right). 

We clearly see that the radial-arc configuration of the two simulated merging fold images~$A$ and $B$ does not resemble the morphology of the observed fold images. 
The parities do not match, i.e.~the simulated image~$A$ has the same parity as the simulated image~$C$, while images~$B$ and $C$ have the same parities in the observed configuration.
This becomes more obvious in the coloured versions in Fig.~\ref{fig:features}, and is even clearer when we compare the observed positions of the bright feature on the SE extremity  of image~$C$, and the corresponding feature in observed images~$A$ and $B$,  with the positions of this feature in simulated images~$A$ and $B$.
Furthermore, an NFW-profile causes a perfect vertical alignment of the three multiple images while image~$C$ in the observations in Fig.\ref{fig:ACS_ABC} has a small offset from the line through images~$A$ and $B$.
Thus, it is unlikely that Hamilton's Object is a radial-arc configuration. 

\begin{figure*} % [h!]
\centering
\includegraphics[width=\textwidth]{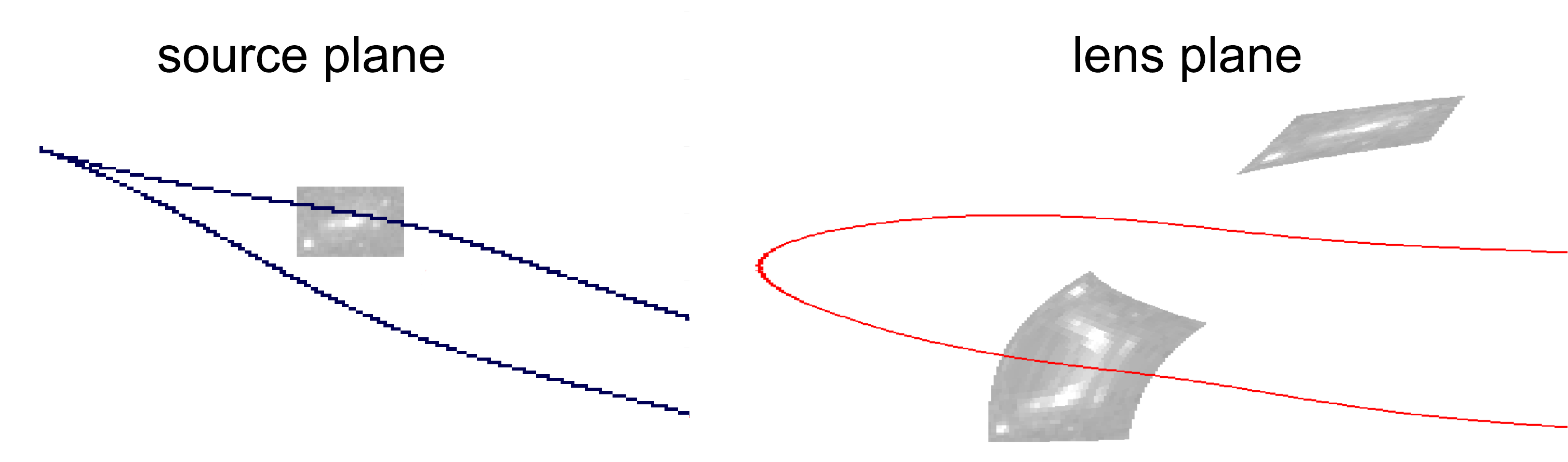}
\caption{Simulation of the multiple-image generation using the \textit{Hera} dark matter profile: same as Fig.~\ref{fig:NFW} but for the mass density profile \textit{Hera} detailed in Meneghetti et al. 2017. The parities of the simulated images match the observations.}
\label{fig:Hera}
\end{figure*}

Next, we replace the NFW profile by a realistically simulated generic cluster lens called \textit{Hera}.
The deflecting mass density profile is a high-resolution zoom-in simulation of an N-body cosmological simulation, as detailed in
Meneghetti et al. (2017). 
 It is located at redshift $z_\mathrm{d}=0.507$, which is close to the lens redshift $z_\mathrm{d}=0.53$ of the galaxy cluster lens discussed here. 
We place image~$C$ as a model for the background source close to the caustic belonging to a source redshift of $z_\mathrm{s}=0.82$, as shown in Fig.~\ref{fig:Hera} (left). 
As a lensing result, we obtain a triple-image configuration shown in Fig.~\ref{fig:Hera} (right), which clearly resembles the observed configuration. 
The parities match and the difference in the relative orientations of the fold images with respect to the critical curve can be explained by the relative orientation of the source and the caustic curve. 
In order to obtain the observed small horizontal offset between images~$A$ and $B$ with respect to image~$C$, the source must lie closer to the caustic line than to the caustic cusp point at the left-most position of the caustic in Fig.~\ref{fig:Hera}. 
As detailed in 
 Schneider, Ehlers and Falco (1992),  (chapter 6), multiple images caused by lensed sources close to the cusp point align almost perfectly on top of each other and often merge into giant arcs, which is not the case here.

\section{Gravitational lens reconstruction}
\label{sec:lens}
\subsection{Applicable lens reconstructions}
\label{sec:lens_intro}

Lens reconstructions aiming at constraining the mass density map in the entire lensing region usually require a high density of multiple lensed images in the cluster region or complementary information about the gravitational lens, such as the positions, surface brightness or velocity dispersion information of the cluster member galaxies, or x-ray emission maps. 
The more data that are available, the more accurate and precise the reconstruction of the mass density profile: see e.g. Johnson and Sharon (2016), Lotz et al. (2017). 

For Hamilton's Object, as detailed in Sections~2 and 3, although we have detailed information on this one triple configuration, the available information on the surrounding galaxy cluster is still very sparse and limited to SDSS photometry.
A gravitational lens reconstruction over the entire lensing region is thus not yet possible due to the lack of further (spectroscopically confirmed) multiple lensed systems.  
We therefore use the observation-based approach detailed in 
Wagner (2019) for a \emph{local} lens reconstruction in the vicinity of the three confirmed multiple images.
The goal is the determination of local lens properties at the positions of the multiple images directly from the observable features.
In this way, we obtain lens properties that do not require any specific assumptions about the global morphology of the mass density profile in the lensing region. 
The formalism is based on Chapter~6 of Schneider, Ehlers and Falco (1992). and exploits the fact that local lens properties can be retrieved from a Taylor expansion around multiple image locations or around points on the critical curve, respectively.
Consequently, these local lens properties can be derived for any gravitational lens and do not require specific assumptions about the global mass density profile.
Section~\ref{sec:prerequisites} details the necessary ingredients and prerequisites to perform the local lens reconstruction.
Subsequently, Sections~\ref{sec:results} and \ref{sec:eval} give an overview of reconstructed local lens properties and their interpretation, respectively.
Lastly, we use the local lens properties in Section~\ref{sec:source} to reconstruct the surface brightness profile of the spiral galaxy which is the common background source of all three observed images.

\subsection{Local lens properties from three resolved multiple images}
\label{sec:prerequisites}

\begin{figure*} %[h!]
\centering
\includegraphics[width=0.6\textwidth]{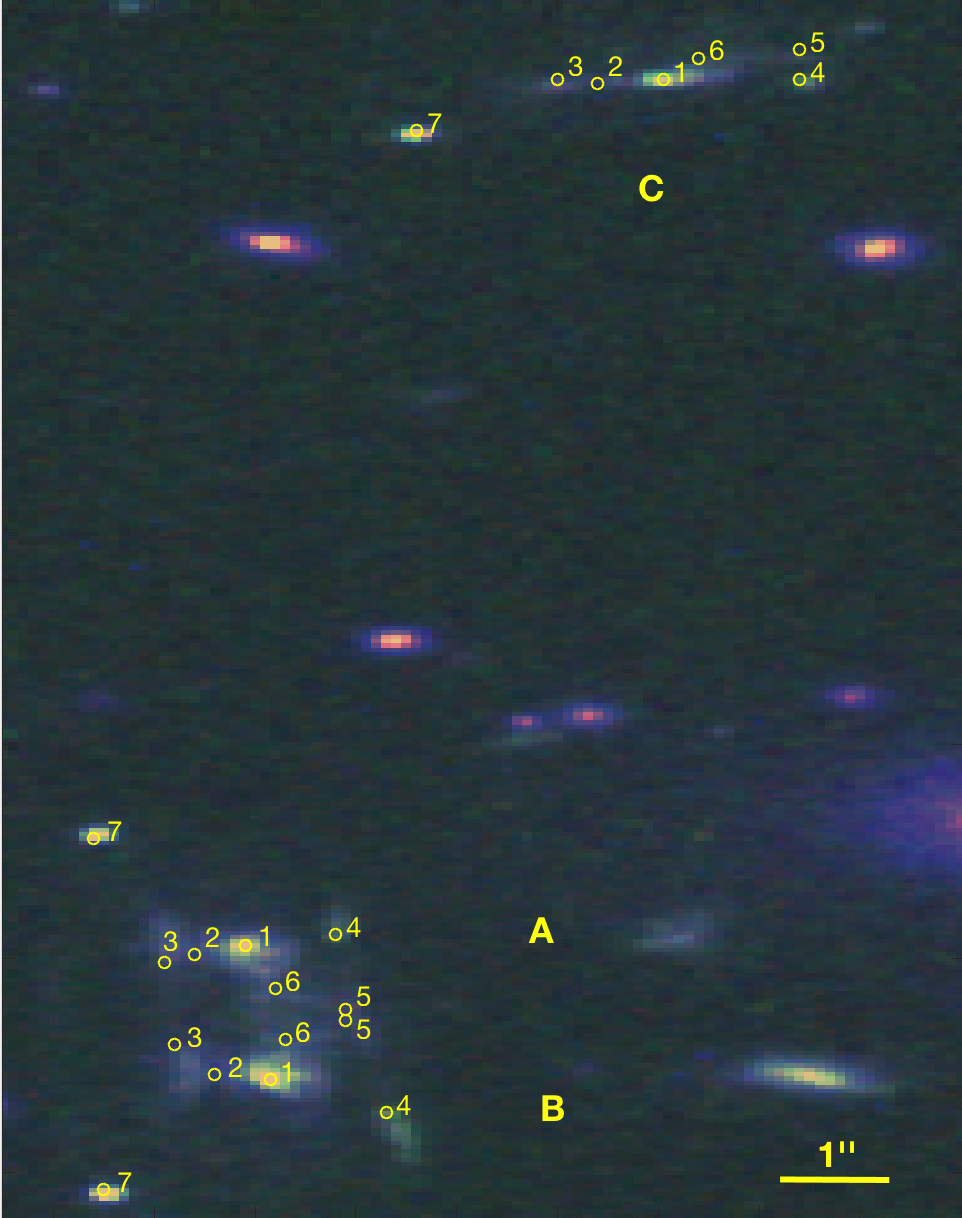}
\caption{Identification of individual surface brightness features in all multiple images in the multi-band HST observations: the two fold images~$A$ and $B$ show seven corresponding features (the positions are marked by the yellow circles annotated by 1-7 in each image) which can also be found in the counter-image~$C$.
The figure is stretched E-W relative to N-S by a factor 2.2 in order to clarify the appearance of the surface brightness features.}
\label{fig:features}
\end{figure*}

Hamilton's Object consists of three multiple images showing resolved features that can be identified in the intensity profiles of the images.
Fig.~\ref{fig:features}
shows the seven identifiable features, after assembling a coloured version of the HST observation consisting of F606W, F814W and the F110W filter bands (longer wavelengths are not used here because of poorer resolution). 
These features have been located to a precision of one pixel, corresponding to 0.09" on the sky at the given pixel resolution of the F110W filter band, which has the lowest resolution of the three filter bands used.

This multiple image configuration with the seven features fulfils the prerequisites to be analysed by the approach summarised in Wagner (2019). 
This means that we assume that the local lens properties extending over the image areas on the sky covered by these seven features can be assumed constant to a good approximation.
Because feature~7 is on the edge of the galaxy (see discussion above in Sec. \ref{sec:spec}),
we decided to investigate its influence on the reconstruction of the local lens properties, and so we first determine the local lens properties excluding feature~7 from the analysis and subsequently repeat the procedure including it. Comparing both reconstructions should reveal whether the local lens properties can be really assumed constant over the extended area covered by all features or whether this approximation is only valid for the smaller area covered by features~1-6.

We use the implementation of  the software package \texttt{ptmatch}\footnote{Available at https://github.com/ntessore/imagemap} for our calculations. 
Details about the method can be found in Wagner (2019), and another example, an analysis of a five-multiple-image configuration in the galaxy cluster CL0024, can be found in Wagner, Liesenborgs and Tessore (2018) . 
The resolved features in the galaxy images in CL0024 were identified in a single filter band observation, 
as were those in Hamilton's Object (Figs. \ref{fig:F110W}, \ref{fig:ACS_ABC} and \ref{fig:C3}), but the features in Hamilton's Object became clearer after stacking the three filter bands mentioned above into a coloured picture, and the combined image was needed in order to measure local lens properties.
(Fig.~\ref{fig:features}).
Due to the extensions of images~$A$ and $B$ orthogonal to the critical curve, Hamilton's Object also provides one of the first lensed galaxies with resolved surface brightness features from which we can reconstruct a local approximation to the critical curve between the images. 

\subsection{\texttt{ptmatch} results}
\label{sec:results}

We first run \texttt{ptmatch} using the six features~1-6 in each multiple image, as marked in Fig.~\ref{fig:features}, assuming uncorrelated imprecisions in $x$- and $y$-directions of 1 pixel. 
Afterwards, we repeat the \texttt{ptmatch} analysis using all seven features in the multiple images.
As the reference image, we choose image~$A$ because it covers a larger area than image~$C$ and it yields tighter confidence bounds for the local lens properties than using image~$B$ as reference. 
The resulting local lens properties are summarised in Table~\ref{tab:llp}.
The $\chi_{\mathrm{red}}^2 = 0.87$ smaller than one, indicates that the transformation of multiple images onto each other can be successfully set up on the basis of these six features. 
Since it is closer to one than the $\chi_\mathrm{red}^2$ when using image~$B$ as reference image, the lens properties are less prone to overfitting.
For all seven features, we obtain $\chi_\mathrm{red}^2 = 1.64$. Hence, there is a slight bias and a possible hint for non-constant local lens properties over the larger area of all seven features.
The local lens properties determined from the transformations are the following (for definitions of these parameters, see Wagner and Tessore (2018)) :
\begin{itemize}
\item The relative ratios between convergences $\kappa$
\begin{equation}
f_i \equiv \dfrac{1-\kappa_A}{1-\kappa_i} \;, \quad i = B, C \;,
\label{eq:f}
\end{equation}
is a measure of the mass density ratio between the different multiple image positions because $\kappa_i$, $i=A, B, C$ is the mass density at the position of a multiple image scaled by the so-called critical mass density to obtain a dimensionless quantity.
\item The reduced shear
\begin{equation}
\boldsymbol{g}_i = (g_{i,1}, g_{i,2}) \equiv \dfrac{\boldsymbol{\gamma}_i}{1-\kappa_i} \;, \quad i=A, B, C \;,
\label{eq:g}
\end{equation}
is a measure of the local distortion per mass density that the lens exerts at the position of a multiple image. From its components, the magnitude and the direction of the reduced shear can be determined as
\begin{equation}
|\boldsymbol{g}_i| = \sqrt{g_{i,1}^2 + g_{i,2}^2} \;, \quad \varphi_{g,i} = \dfrac{1}{2} \text{atan} \left( \dfrac{g_{i,2}}{g_{i,1}} \right) \;,
\label{eq:g_pol}
\end{equation} 
for $ i=A,B, C$.
At a critical curve, $|\boldsymbol{g}|=1$, with values larger than 1 inside and decreasing values smaller than 1 outside. 
\item The magnification ratio 
\begin{equation}
\mathcal{J}_i \equiv \dfrac{\text{det}(M_A)}{\text{det}(M_i)} \;, %\quad
M_i = (1-\kappa_i) \left(\begin{matrix} 1-g_{i,1} & -g_{i,2} \\ -g_{i,2} & 1+g_{i,1} \end{matrix} \right)  \;
\label{eq:J}
\end{equation}
between images~$i=B$ and $C$ to image~$A$ is determined from the $f_i$ and $\boldsymbol{g}_i$. It serves as a cross-check quantity, if flux ratios between the multiple images are measured. 
For images in a fold configuration, we expect a magnification ratio close to -1. 
This is physically interpreted as having equal magnifications for both images but a reversed parity.
Thus, $\mathcal{J}$ shows the relative parity between the multiple images. 
If local lens properties with relative parities are retrieved that are inconsistent with the observed relative parities, the respective \texttt{ptmatch} result can be refuted and it can be doubted whether the chosen multiple image candidates actually belong to the same background source.  
\end{itemize}

\begin{table*}[t]
 \caption{Synopsis of local lens properties obtained from Hamilton's Object using features~1-6 (left) and using all 7 features marked in Fig.~\ref{fig:features} (right): name of the local lens property (first column), its most-likely value, ML (second column), its mean value, Mean (third column), and the width of the sampled distribution between the 16th and the 84th percentile, Std (fourth column), based on 10 000 samples around the positions of the features marked in Fig.~\ref{fig:features}. Details concerning the treatment of the statistics can be found in Wagner and Tessore (2018) and Wagner, Liesenborgs and Tessore (2018)}
\label{tab:llp}
\begin{center}
\begin{tabular}{c|rrr}
\hline
%\noalign{\smallskip}
Lens prop.         &	ML	&   Mean	&	Std	\\
%\noalign{\smallskip}																									
\hline																									
%\noalign{\smallskip}																									
$g_{A,1}$	&	-1.78&	-2.02&	0.73	\\
$g_{A,2}$	&	-0.19	&	-0.20&	0.08	\\
%\noalign{\smallskip}																									
\hline																									
%\noalign{\smallskip}																									
$\mathcal{J}_B$	& -1.27 	& -1.27	& 0.10\\
$f_B$	& 0.59	& 0.60	& 0.24 \\
$g_{B,1}$	& -0.56	& -0.56	& 0.21 \\
$g_{B,2}$	& -0.25	& -0.25	& 0.05 \\
%\noalign{\smallskip}																									
\hline																									
%\noalign{\smallskip}																									
$\mathcal{J}_C$	& -0.58 &	-0.58 &	0.10	\\
$f_C$	&  0.51	&	0.59	&	2.03	\\
$g_{C,1}$	& -0.03	&	-0.11	&	2.14 \\
$g_{C,2}$	&	0.04	&	0.02	&	0.43	\\
%\noalign{\smallskip}																									
\hline																									
\end{tabular}
\hspace{5ex}
\begin{tabular}{c|rrr}
\hline
%\noalign{\smallskip}
Lens prop.         &	ML	&   Mean	&	Std	\\
%\noalign{\smallskip}																									
\hline																									
%\noalign{\smallskip}																									
$g_{A,1}$	&	-1.97&	-2.10&	0.53	\\
$g_{A,2}$	&	-0.03	&	-0.04&	0.06	\\
%\noalign{\smallskip}																									
\hline																									
%\noalign{\smallskip}																									
$\mathcal{J}_B$	& -1.37 	& -1.36	& 0.08\\
$f_B$	& 0.58	& 0.57	& 0.15 \\
$g_{B,1}$	& -0.53	& -0.53	& 0.14 \\
$g_{B,2}$	& -0.14	& -0.14	& 0.04 \\
%\noalign{\smallskip}																									
\hline																									
%\noalign{\smallskip}																									
$\mathcal{J}_C$	& -0.53 &	-0.53&	0.07	\\
$f_C$	&  0.42	&	0.45	&	0.42 \\
$g_{C,1}$	& 0.09	&	0.07	&	0.45 \\
$g_{C,2}$	&	0.13	&	0.12	&	0.06	\\
%\noalign{\smallskip}																				
\hline																								
\end{tabular}
\end{center}
\end{table*}

Comparing the most-likely value and the mean value for each local lens property in Tab.~\ref{tab:llp} (left) and (right) individually, we find that each of the ML and mean values agree with each other within the 68\% confidence level. 
There are three confidence bounds that exceed the mean and most-likely lens property value by about an order of magnitude for image~$C$ in Tab.~\ref{tab:llp} (left), while only one local lens property in image~$C$ shows such behaviour in Tab.~\ref{tab:llp} (right).
However, most of these local lens properties, the components of the reduced shear, are close to zero and confidence bounds larger than the value can occur and also did occur in the example of CL0024. 
In addition, we observe a reduced size of the confidence bounds when increasing the image area, as Std in Tab.~\ref{tab:llp} (left) is always larger than Std in Tab.~\ref{tab:llp} (right).
Beyond the relatively good quality of confidence bounds, there is a high degree of coincidence between the most-likely and mean local lens property value, corroborating the successful pairwise transformation of the multiple images onto each other and thus, yielding further evidence that these three images originate from the same background source. 
Comparing the most-likely local lens properties of Tab.~\ref{tab:llp} (left) and Tab.~\ref{tab:llp} (right) with each other, we find coincidences within the confidence bounds for all local lens properties but $g_{A,2}$ and $g_{B,2}$. 
This high degree of coincidence between the two \texttt{ptmatch} reconstructions implies that the bias when using all 7 features is really very small.

\subsection{\texttt{ptmatch} evaluation}
\label{sec:eval}

Comparing the relative positions of the six (or seven) features between the multiple images, we find that images~$B$ and $C$ have the same parity and image~$A$ has reversed parity. 
These relative parities are typical for tangential cusp configurations of three multiple images and are also found for the analogous configuration in CL0024.
Hence, \texttt{ptmatch} yields the correct relative parities with $\text{sign}(\mathcal{J}_B) = \text{sign}(\mathcal{J}_C) = - \text{sign}(\mathcal{J}_A)$.
In addition, $\mathcal{J}_B$ is close to -1, as expected for the fold configuration consisting of images~$A$ and $B$. 
$\mathcal{J}_B$ is smaller than -1, indicating that image~$B$ is more strongly magnified than image~$A$. 
Comparing the predicted magnification ratios of Tab.~\ref{tab:llp} (left) and (right) to the observed flux ratios for all available wavelengths in Tab.~\ref{tab:mags}, we find an overall good degree of agreement. 
%% ?????(REG)  
As Tab.~1 only contains the flux ratios of the central flux from core feature~1, we can conclude that the eastern and western additional parts of the galaxy only add minor contributions to the surface brightness distribution, which seems realistic from visual inspection of Fig.~\ref{fig:features} and confirms the consistency of our analysis with the observations.

\begin{figure*} % [h!]
\centering
\includegraphics[width=0.6\textwidth]{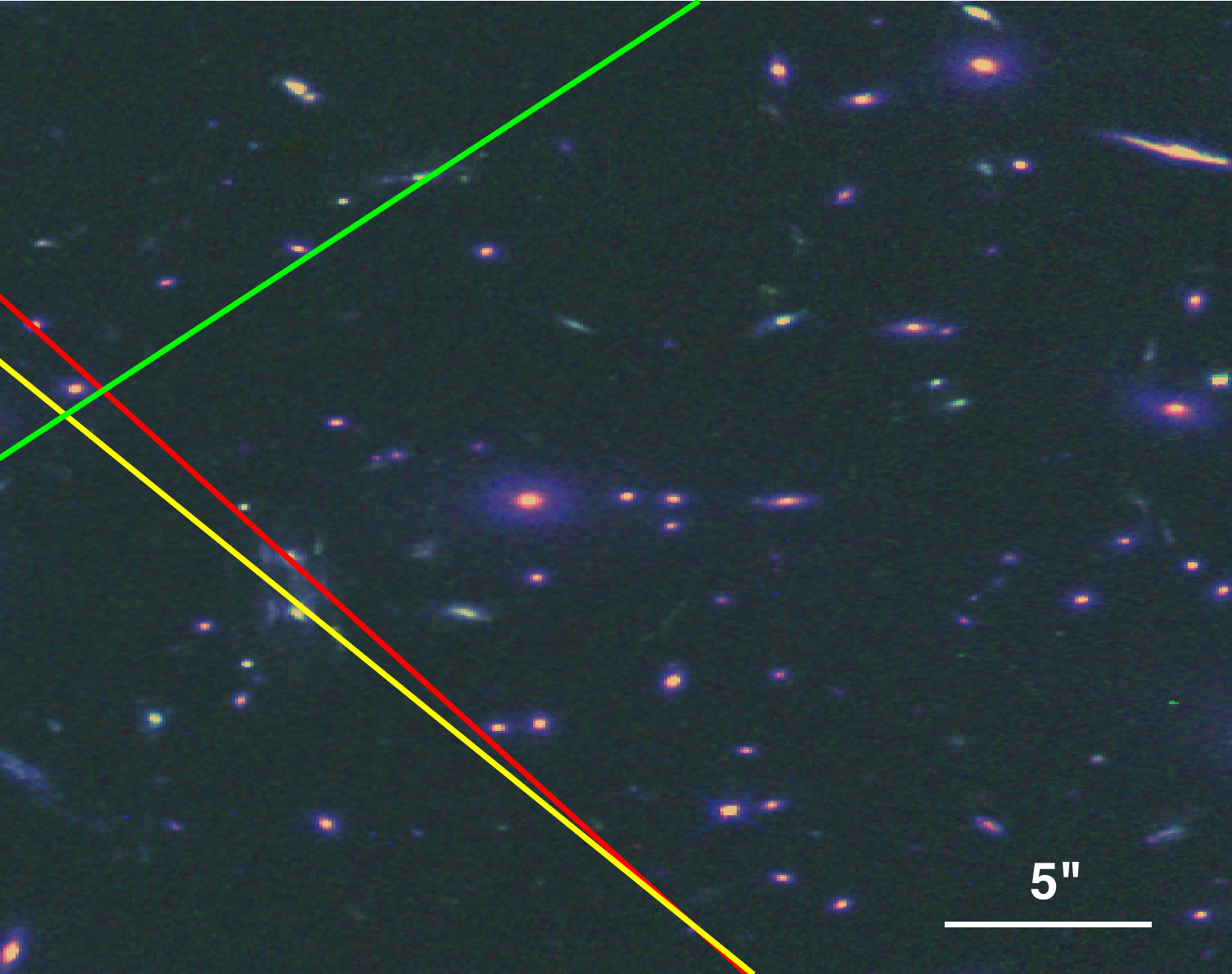}
\caption{Directions of reduced shears for image~$A$ (red line), image~$B$ (yellow line), and image~$C$ (green line) starting at the anchor point of each multiple image as obtained from Tab.~\ref{tab:llp} (left). The crossings give a rough estimate of the extension of the critical curve for the source redshift of 0.82.
The figure is stretched E-W relative to N-S by a factor of 2 in order to clarify the appearance of the surface brightness features}
\label{fig:apex}
\end{figure*}

Next, we evaluate the $f_i$, which follows the same trend in Tab.~\ref{tab:llp} (left) and (right).
According to Eq.~\eqref{eq:f}, the positivity of both $f_i$ implies that all convergences are either smaller or larger than one. 
Since the multiple images are supposed to form a tangential cusp configuration at an outer critical curve, it is more likely that all convergences are smaller than one. 
In that case, $1 > f_B > f_C$ implies that $\kappa_A > \kappa_B > \kappa_C$.
So we can conclude that the mass density at the image positions is decreasing outwards, if image~$A$ is closer to the cluster centre than image~$B$, which, in turn, is closer to the cluster centre than image~$C$. 

The interpretations of $\mathcal{J}_i$ and $f_i$ are consistent with the reduced shears. 
Using Eq.~\eqref{eq:g_pol} to convert the reduced shear components to their polar representation, we find that $|\boldsymbol{g}_A| = 1.79 > |\boldsymbol{g}_B| = 0.62$ for Tab.~\ref{tab:llp} (left) and $|\boldsymbol{g}_A| = 1.97 > |\boldsymbol{g}_B| = 0.55$ for Tab.~\ref{tab:llp} (right). 
This implies that image~$A$ is closer to the cluster centre and farther from the critical curve which occurs at $|\boldsymbol{g}|=1$. 
Consequently, we expect image~$B$ to be more strongly magnified due to its position closer to the critical curve. 

\begin{figure*}% [h!]
\centering
\includegraphics[width=0.49\textwidth]{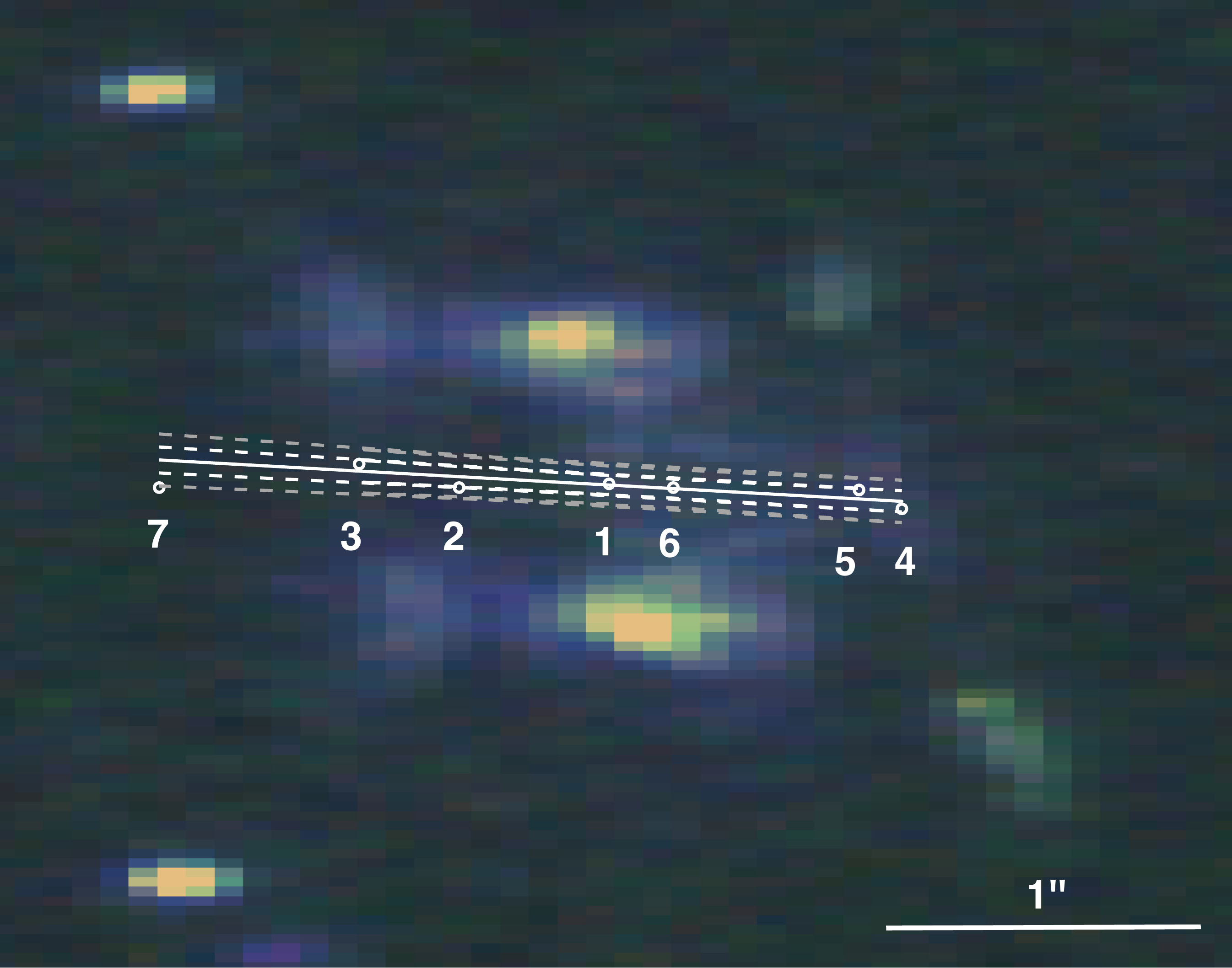}
\hfill
\includegraphics[width=0.49\textwidth]{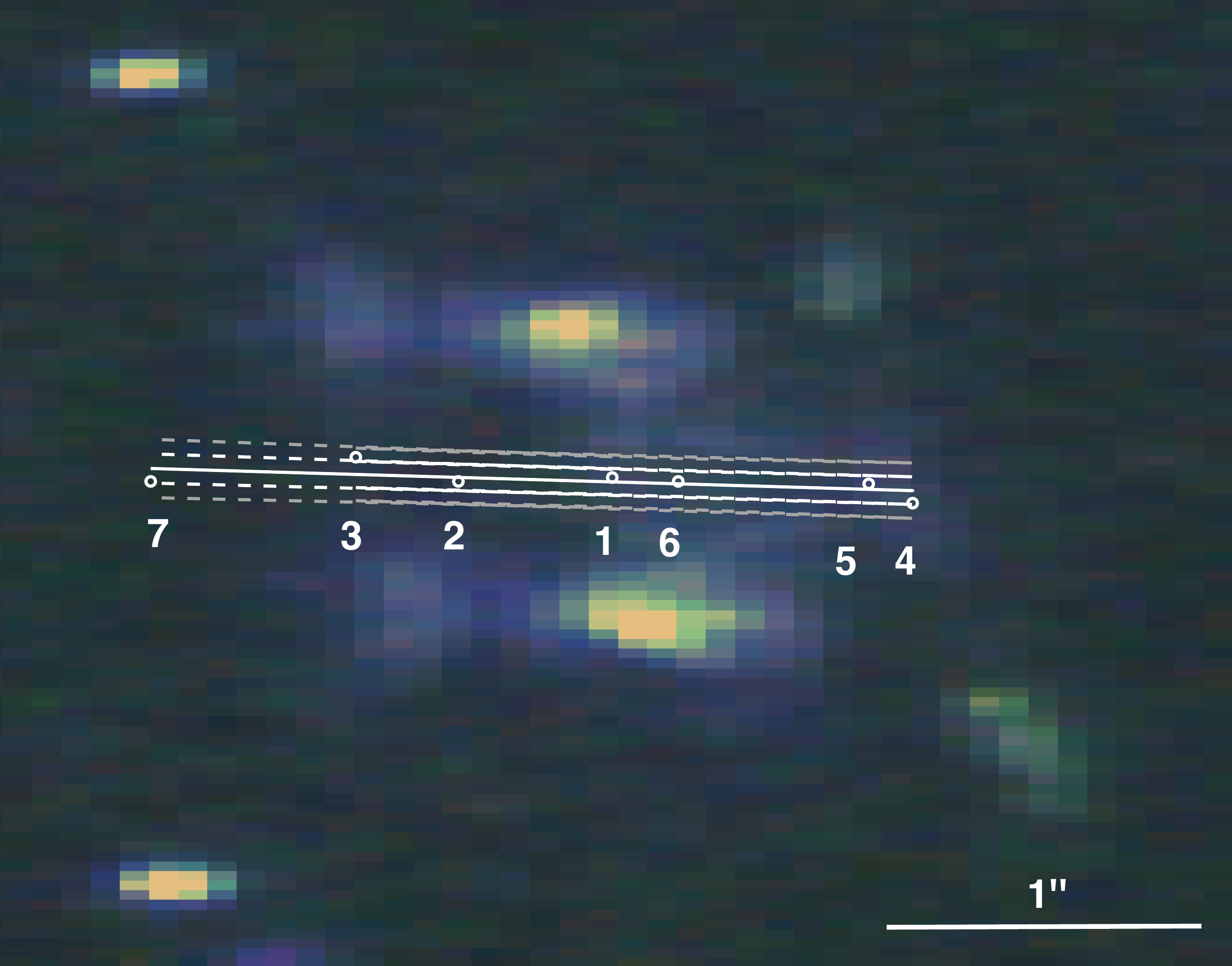}
\caption{Reconstruction of the critical curve between images~$A$ and $B$ based on features~1--6 (left) and based on all seven features (right): all points on the critical curve determined from the respective (white circles) are fitted to a linear approximation of the critical curve (solid white line) and its confidence bounds given at 68\% confidence level (dashed white line) and at 95\% confidence level (dashed grey line).
The figure is stretched E-W by a factor of 3 relative to N-S in order to clarify the appearance of the surface brightness features.}
\label{fig:cc}
\end{figure*}

In Wagner (2017) and Wagner and Tessore (2018), the peak position of a parabolic approximation to the critical curve was derived based on the positions and shapes of the three cusp images. 
While the multiple images of Hamilton's Object are too far from the peak position to allow for a parabolic approximation to the critical curve, we roughly estimate the peak position by plotting the most-likely directions of the reduced shears for all images, starting at the central position of each image. 
The latter is determined by \texttt{ptmatch} as the so-called anchor point of each image. 
(For image~$A$, it is fixed to the centroid of all feature positions.)
The crossings of the three reduced shear directions of Tab.~\ref{tab:llp} (left), as shown in Fig.~\ref{fig:apex} give a first impression of the extent of the critical curve for the redshift of the source. 
In this analysis, we first note the bias of the reconstruction in Tab.~\ref{tab:llp} (right) because the direction derived for image~$A$ crosses the one from Tab.~\ref{tab:llp} (left) due to the different sign of $g_{A,2}$. Yet, within the confidence bounds, negative $g_{A,2}$ and shear directions pointing to the most-likely cusp position of the critical curve can still occur.

Without a fourth counter-image on the opposite side of the cluster centre, we cannot constrain the overall projected mass in the area on the sky enclosed by the multiple images. 
The object at RA 22h 30m 07.25s and dec. -08d 09m 18.17s 
%$\alpha=337.5302216^\circ$, $\delta=-8.155048891^\circ$
is a potential candidate for the fourth image but has not been measured spectroscopically because it fell outside
the KCWI field of view.
% It could not be included in the \texttt{ptmatch} analysis as the surface brightness features could not be identified with certainty. %due to the low resolution.
Lastly, we reconstruct the critical curve between images~$A$ and $B$ from the features marked in Fig.~\ref{fig:features}. 
We first apply the approach outlined in Wagner (2017) for all matching features 1--6 and subsequently for all seven features, such that the position of the critical curve is given as the mean position between the positions of corresponding features in the two images. 
Fig.~\ref{fig:cc} (left) shows the resulting positions as white circles. 
Using the features 1--6 to determine local lens properties across the entire area covered by the multiple image on the sky, we implicitly assume that these lens properties hardly change over the area enclosed by the features.
Thus, the points can be fitted by a line and thereby approximate the critical curve between the images, as the solid white line in Fig.~\ref{fig:cc} shows. 
The approximation has quite tight confidence bounds, indicated by the 68\% and 95\% confidence bounds marked by the dashed white and dashed grey lines, respectively. 
The linear approximation to the critical curve is also extended to investigate how well the estimate of the critical curve position determined from feature~7 matches  the linear approximation set up by features~1--6. 
We read off Fig.~\ref{fig:cc} (left) that is still at the boundary of the 95\% confidence bound. 
The linear approximation to the critical curve based on all seven features is shown in Fig.~\ref{fig:cc} (right) for comparison. While the residual norm per feature point pair from image~$A$ and $B$ for the former reconstruction is 0.46, it amounts to 0.54 for this approximation. 
Hence, the fit only deteriorates slightly.

% An attempt to extract local lens properties on the scale of the individual features, for instance by determining the quadrupole moments of surface brightness for the central feature 1, fails due to insufficient resolution at the redshift of the source.  %%%%%%%%%%%%%%%%%%%%%%

We conclude that the dark matter distribution is homogeneous on arc-second length scales (6 kpc) at the positions of the multiple images, otherwise \texttt{ptmatch} could not have found such a consistent solution. %The nuclear bulges have bright features about $0^{''}.2$ arcecs. west (Fig.\ref{ACS_ABC}), which are also preserved 
%in the \text{ptmatch} translations  (????) , and this indicates that the DM distribution is homogeneous on even smaller scales ($\sim 1$~ kpc). 
Gradients in the dark matter densities, originating from small-scale dark matter inhomogeneities on sub-arcsec. scales cannot be resolved at present. %%% Nuclei ???? 
Andrade et al. (2020) have used strongly lensed images in eight galaxy clusters to measure dark matter density profiles over the scale of 10 kpc to 100 kpc.

\subsection{Source reconstruction}
\label{sec:source}

\begin{figure*}. %. [h!]
\centering
\includegraphics[width=0.325\textwidth]{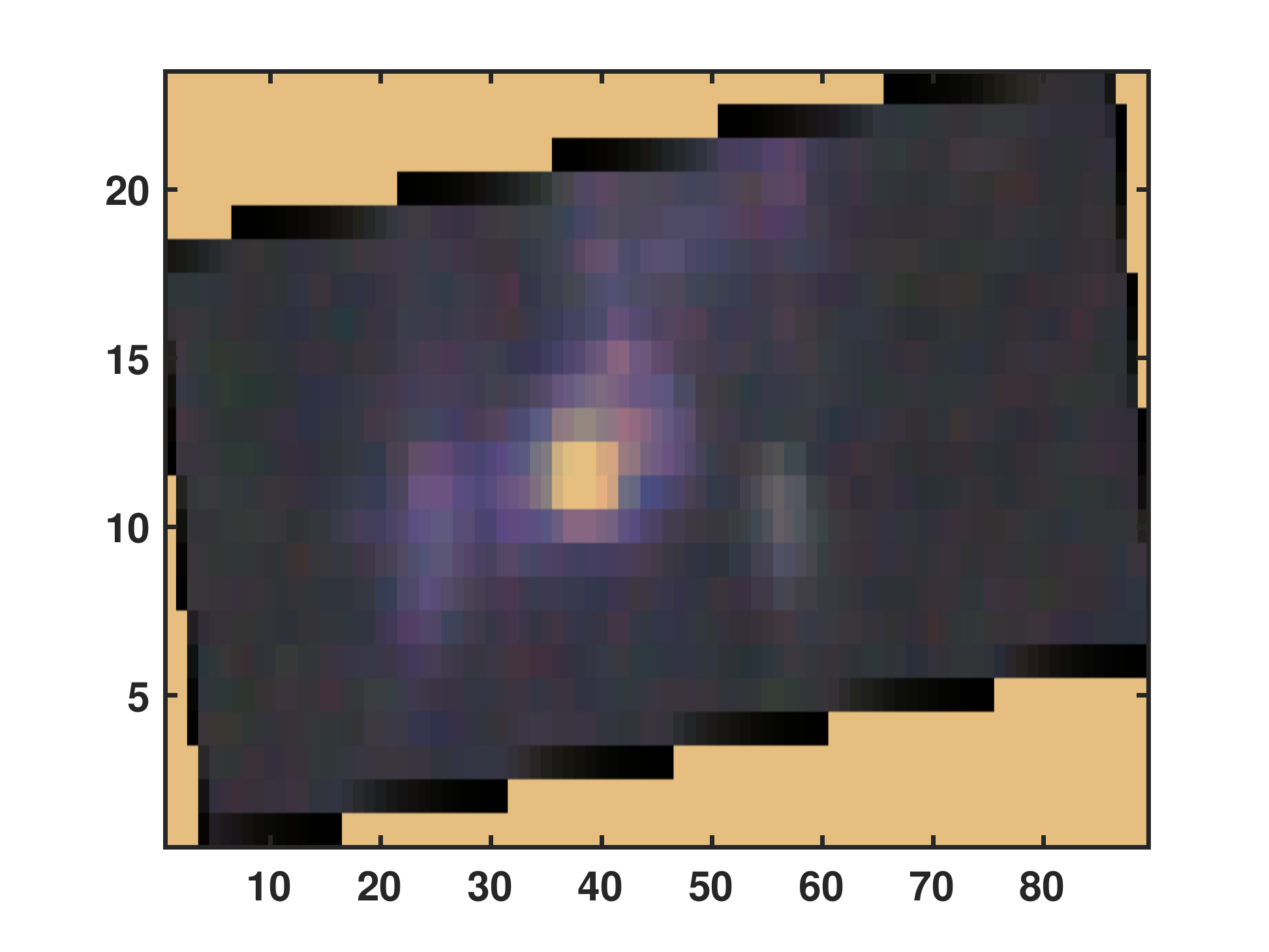}
\includegraphics[width=0.325\textwidth]{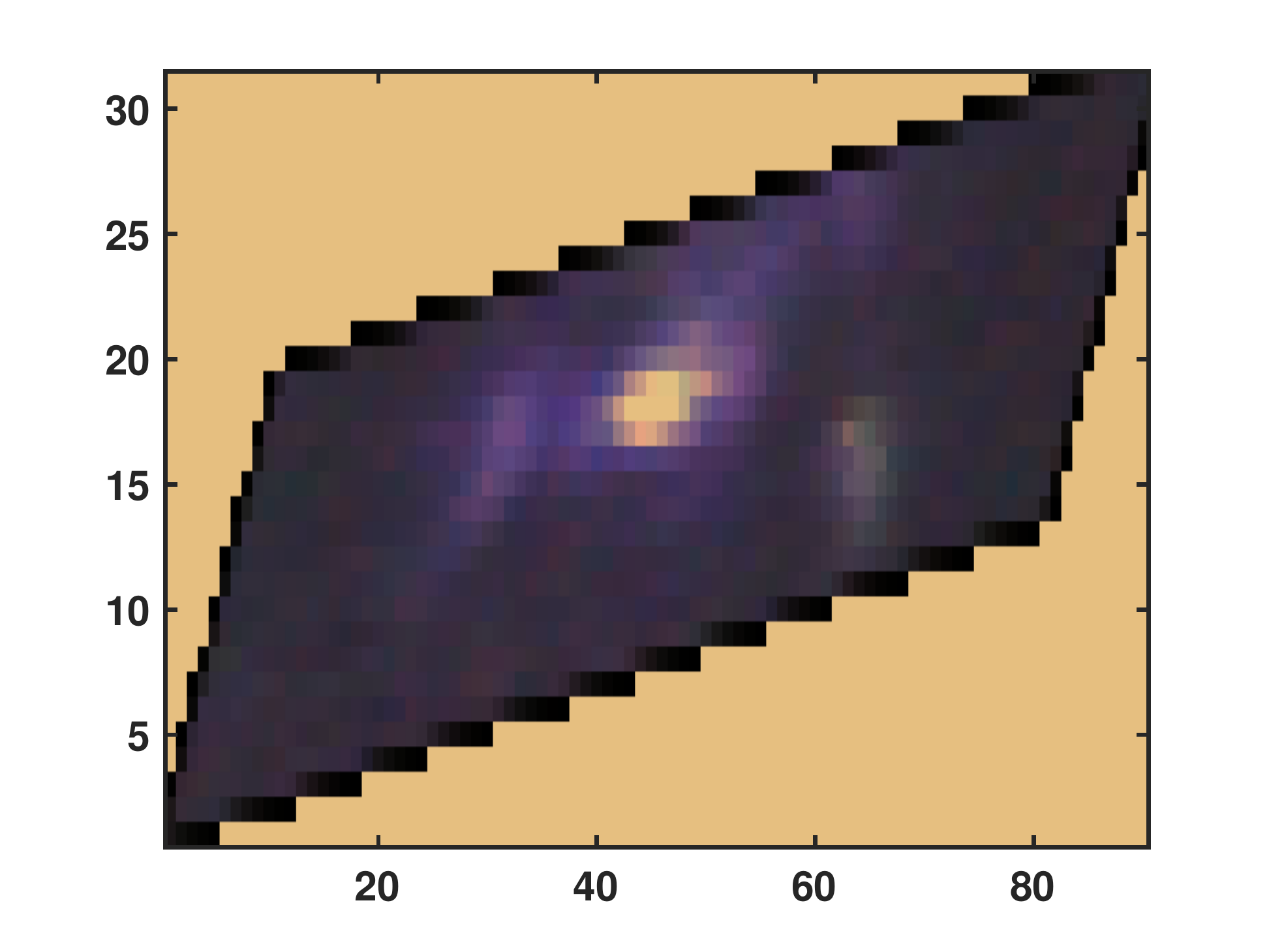}
\includegraphics[width=0.325\textwidth]{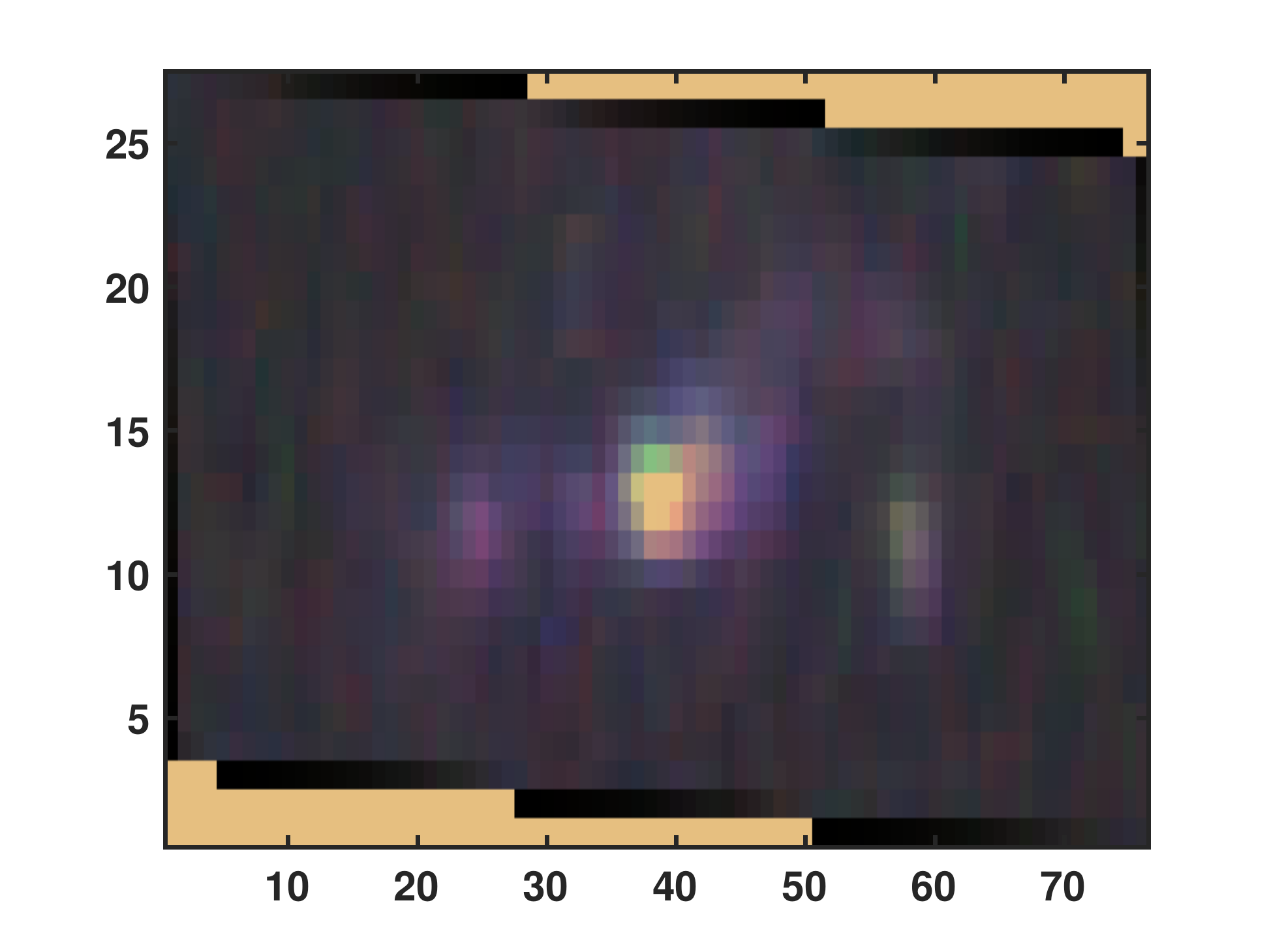}
\caption{Back-projections of each multiple image~$A$ to $C$ (from left to right) to the source plane to obtain a reconstruction of the common source based on the local lens properties of Tab.~\ref{tab:llp} (left).
%  The images are stretched E-W relative to N-S. The scale unit of '10' corresponds with  XXX    arcsecs.
}
\label{fig:source}
\end{figure*}

In order to further cross-check the resulting local lens properties, we use the software package \texttt{srcim}\footnote{Also available at https://github.com/ntessore/imagemap} to back-project each multiple image back to the source plane by means of its magnification matrix $M_i$ (see Eq.~\eqref{eq:J}).
As the approach only constrains the ratios of convergences, the source can only be reconstructed up to an overall scale factor. 
The latter is chosen such that $M_A$ has unit determinant. 
Using the local lens properties of Tab.~\ref{tab:llp} (left), we arrive at the three source reconstructions shown in Fig.~\ref{fig:source}. 
The reconstuctions based on Tab.~\ref{tab:llp} (right) look very similar and are omitted, as the fit to the observations was slightly worse and biased compared to the local lens properties of Tab.~\ref{tab:llp} (left).

All three back-projections show a high degree of coincidence from visual comparison. 
A detailed astrophysical analysis of the source object is impeded by the low resolution of the reconstruction, which is based on the pixel-wise back-projection of the image data and does not make any additional assumptions about the morphology or the degree of smoothness of the source surface brightness profile. 
Hence, the reconstructions in Fig.~\ref{fig:source} represent the information on arc-second scales contained in the surface brightness profiles of the multiple images that is common to all approaches with refined, additional assumptions.
 
%%%%%%%%%%%%%%%%%%%%%%
%%%%%%%%%%%%%%
\section{A serendipitous Lyman~$\alpha$ blob at z = 3.2}
\label{sec:blob}

In the KCWI spectral datacube, a strong Lyman~$\alpha$ emission feature was found ~8.2 arcseconds to the west of the main lensed images
A and B, at RA 22h 30m 09.03s, dec -08d 09m 42.7s. The single emission-line feature, at an observed wavelength of 5106  \AA, is identified as a Lyman~$\alpha$ blob (LAB)
at redshift 3.199$\pm0.001$, following the original discovery of such objects by Steidel et al. (2000) (see also Francis et al. 1996) and their subsequent  detections in data from instruments such as the 
Subaru Prime Focus Camera (Matsuda et al 2004) % KCWI (refs....)
 and MUSE (Vanzella et al. 2016). The Lyman~$\alpha$ blob has dimensions of 4 arcsecs N-S and 3 arcsecs. E-W, i.e. 30 x 23 kpc, comparable with similar objects found by Matsuda et al. (2004).
 The observed  Lyman~$\alpha$ flux is $ \sim 1.5 \times 10^{-16}$ ergs cm$^{-2}$ s$^{-1}$ as measured in the KCWI, and at the 
observed redshift of 3.2, this would imply a luminosity of  $1.3 \times 10^{43}$ ergs s$^{-1}$. The object may be magnified by the gravitational
potential of the galaxy cluster, however.  In common with other LAB's, we assume that this object is a gaseous hydrogen halo surrounding
a young cluster of starbursting galaxies and/or AGN (Geach et al. 2016), although the interpretations for LAB's are split between hyperwind activity
from starburst galaxies and inflowing gas (see e.g. Ao et al. (2017, 2020)). The LAB is not detected in any of the  HST images to limiting magnitudes
 of ~27 in F606W and 26 in F160W.

Further observations of this and other LAB's are required.

\section{Conclusions} 
\label{sec:conc}

In a HST observation of an x-ray selected AGN, we have serendipitously discovered an unusual gravitationally lensed image configuration which, for the source redshift, 
is interpreted as a fold image configuration straddling a critical curve caused by a foreground cluster of galaxies, where the cluster mass is about $5 \times 10^{14}$ M$_\odot$, 
and the cluster contains about 60 $L^*$ galaxies, far less than some of the clusters previously characterized via lensing such as those in the CLASH and Hubble Frontiers Fields.
%***  [JG: actually, it's the foreground cluster of galaxies causing the light bundles from the background galaxy to be folded into multiple images.]
The redMaPPer cluster, at a redshift of 0.526, is identified with the faint ROSAT x-ray source 1RXS J222956.9-080823. % (eMACS J2229.9 -0808).
In addition to the fold pair of images, we identify a candidate third or counterimage as a large clumpy disk galaxy at a redshift of 0.8200.  The lensed system is very similar to  system \#12 of Caminha et al. (2017)
%We have used the software package `lenstronomy' to show that an object with the source morphology can be lensed into the configuration of the observed double images when the source is placed just inside the astroidal caustic of the cluster of galaxies, on a `fold' line between the apices on the semi-major and semi-minor axes.

We have followed the observation-based analysis of Wagner \& Tessore (2017) and Wagner (2019)
%Wagner and Bartelmann (2016)
 to constrain the properties of the cluster lens. %the position of the source on the cluster caustic.
The dark matter distribution is homogeneous on arc-second length scales at the positions of the multiple images, otherwise the \texttt{ptmatch} software could not have found such a consistent solution. 
Gradients in the dark matter density, originating from small-scale dark matter inhomogeneities on sub-arcsec scales ($<6$ kpc) cannot be resolved. 

% and to show consistency between the inferred cluster mass and that estimated 
% from the tentative x-ray luminosity (????).  
Such gravitationally folded galaxy images, straddling a critical curve caused by a galaxy cluster, are extremely valuable for investigations of the clumping of dark matter via microlensing of individual supergiant stars or asymmetries in the surface brightness distributions of lensed components. Individual Population III stars and their stellar-mass
black hole accretion disks may possibly be studied using JWST through cluster caustic transits (Windhorst et al. 2019).

% \fixme{Maybe the following sentences are also suitable for the conclusion?}
Further observations are needed to identify other lensed images caused by the cluster and to further characterize the cluster.
Monitoring with future large telescopes could be very fruitful.
%*** JG Maybe we can further add a first rough mass estimate for the cluster and characterise the cluster a bit more compared to other lensing clusters to make it appealing for further studies? ..and I would add that we can also identify more multiple images from other background sources for follow up studies of the entire cluster morphology. 

\section*{Acknowledgements}

This research is based on observations made with the NASA/ESA Hubble Space Telescope obtained from the Space Telescope Science Institute, which is operated by the Association of Universities for Research in Astronomy, Inc., under NASA contract NAS5-26555. The original HST observations are associated with program GO-13305. 
Observations have also been obtained from the Hubble Legacy Archive, which is a collaboration between the Space Telescope Science Institute (STScI/NASA), the Space Telescope European Coordinating Facility (ST-ECF/ESAC/ESA) and the Canadian Astronomy Data Centre (CADC/NRC/CSA). In particular, data were obtained 
from HST programs GO-13671 and GO-14098. Support for Program GO-13305 was provided by NASA through a grant from the Space Telescope Science Institute.

A crucial part of the data presented herein were obtained at the W. M. Keck Observatory, which is operated as a scientific partnership between the California Institute of Technology, the University of California and the National Aeronautics and Space Administration. The Observatory was made possible by the generous financial support of the W. M. Keck Foundation. Observations presented here have included results from the KCWI and NIRC2 instruments, under programs H305 and H315, part of the observing time
allotment made by the University of Hawaii system (and the Institute of Astronomy) to U Hawaii, Hilo for educational and research training purposes following the 
decommissioning and removal of the UH Hilo teaching telescope from the summit of Mauna Kea. We greatly appreciate the help of Luca Rizzi with the calibrations and observations.  Data from the KCWI were analyzed using the KCWI Data Reduction Pipeline.

The results presented here are also partly based on observations obtained at the international Gemini Observatory, a program of the NSF NOIRLab, which is managed by the Association of Universities for Research in Astronomy (AURA) under a cooperative agreement with the National Science Foundation on behalf of the Gemini Observatory partnership.
% : the National Science Foundation (United States), National Research Council (Canada), Agencia Nacional de Investigación y Desarrollo (Chile), Ministerio de Ciencia, Tecnología e Innovación (Argentina), Ministério da Ciência, Tecnologia, Inovações e Comunicações (Brazil), and Korea Astronomy and Space Science Institute (Republic of Korea).
Gemini data were obtained in 2017 and 2018 under programme number GN-2017A-Q-48 which was also part of the observing time
allotment made by the University of Hawaii system (and the Institute of Astronomy) to U Hawaii, Hilo for educational and research training purposes following the 
decommissioning and removal of the UH Hilo teaching telescope from the summit of Mauna Kea.  We acknowledge the help of Andrew Stephens with the queue-scheduled observations.
Gemini data were analyzed using the GMOS-N and the NIRC-2 data reduction and analysis pipelines, operated within IDL.
 
 We have made use of the following databases: the Mikulski Archive
 for Space Telescopes (MAST); the Sloan Digital Sky Survey (SDSS), Data Release 12;
 the Pan-STARRS survey archived at the Hubble Legacy Archive, the
 NASA/IPAC Extragalactic Database (NED), and NASA's Astrophysics Data System Abstract Service (ADS).
 
This research has also made use of data and/or software provided by the High Energy Astrophysics Science Archive Research Center (HEASARC), which is a service of the Astrophysics Science Division at NASA/GSFC.

All spectra presented here were produced using SAOImage DS9, for which development has been made possible by funding from NASA's
Applied Information Systems Research Program, Chandra X-ray Science Center (CXC), and the High Energy Astrophysics Science Archive Center (HEASARC).
Additional funding was provided by the James Webb Space Telescope Mission office at Space Telescope Science Institute to improve capabilities for 3D data visualization.
 
The Pan-STARRS1 Surveys (PS1) and the PS1 public science archive have been made possible through contributions by the Institute for Astronomy, the University of Hawaii, the Pan-STARRS Project Office, the Max-Planck Society and its participating institutes, the Max Planck Institute for Astronomy, Heidelberg and the Max Planck Institute for Extraterrestrial Physics, Garching, The Johns Hopkins University, Durham University, the University of Edinburgh, the Queen's University Belfast, the Harvard-Smithsonian Center for Astrophysics, the Las Cumbres Observatory Global Telescope Network Incorporated, the National Central University of Taiwan, the Space Telescope Science Institute, the National Aeronautics and Space Administration under Grant No. NNX08AR22G issued through the Planetary Science Division of the NASA Science Mission Directorate, the National Science Foundation Grant No. AST-1238877, the University of Maryland, Eotvos Lorand University (ELTE), the Los Alamos National Laboratory, and the Gordon and Betty Moore Foundation.
 
Funding for the Sloan Digital Sky  Survey IV has been provided by the Alfred P. Sloan Foundation, the U.S. Department of Energy Office of Science, and the Participating 
Institutions. 
SDSS-IV acknowledges support and resources from the Center for High Performance Computing  at the University of Utah. The SDSS website is www.sdss.org.
% \paragraph{
SDSS-IV is managed by the Astrophysical Research Consortium for the Participating Institutions of the SDSS Collaboration (q.v.).
% including the Brazilian Participation Group, the Carnegie Institution for Science, Carnegie Mellon University, Center for Astrophysics | Harvard \& Smithsonian, the Chilean Participation 
%Group, the French Participation Group, Instituto de Astrof\'isica de Canarias, The Johns Hopkins University, Kavli Institute for the Physics and Mathematics of the 
%Universe (IPMU) / University of Tokyo, the Korean Participation Group, Lawrence Berkeley National Laboratory, Leibniz Institut f\"ur Astrophysik Potsdam (AIP),  Max-Planck-Institut f\"ur Astronomie (MPIA Heidelberg), Max-Planck-Institut f\"ur Astrophysik (MPA Garching), Max-Planck-Institut f\"ur Extraterrestrische Physik (MPE), 
%National Astronomical Observatories of China, New Mexico State University, New York University, University of Notre Dame, Observat\'ario Nacional / MCTI, The Ohio State 
%University, Pennsylvania State University, Shanghai Astronomical Observatory, United Kingdom Participation Group, Universidad Nacional Aut\'onoma 
%de M\'exico, University of Arizona, University of Colorado Boulder, University of Oxford, University of Portsmouth, University of Utah, University of Virginia, University 
%of Washington, University of Wisconsin, Vanderbilt University, and Yale University.

We would like to thank Nicolas Tessore for his open source lens reconstruction programs and Adi Zitrin and Liliya Williams for useful comments. JW gratefully acknowledges the support by the Deutsche Forschungsgemeinschaft (DFG) WA3547/1- 3.
We thank the Bologna Lens Factory, and in
particular Massimo Meneghetti, for providing us with the \textit{Hera} simulation.

 REG acknowledges receipt of a grant from the NASA Astrophysics Data Analysis Program (NX15AE61G) to the Research Corporation of the University of Hawaii  from 2015 to 2018,
 when the initial stages of this work were performed. 
 Po-Chieh Huang acknowledges support for his undergraduate research work at the University of Hawaii, Hilo,
 while on an exchange studentship from Chung Yuan Christian University, Taiwan.

The authors wish to recognize and acknowledge the very significant cultural role and reverence that the summit of Maunakea has always had within the indigenous Hawaiian community.  We are most fortunate to have the opportunity to conduct observations from this mountain, using both the Gemini-North and Keck telescopes.

%%%%%%%%%%%%%%%%%%%%%%%%%%%%%%%%%%%%%%%%%%%%%%%%%%

%%%%%%%%%%%%%%%%%%%% REFERENCES %%%%%%%%%%%%%%%%%%

\section*{Data availability}

The HST and Pan-STARRS PS1 data used here are available from the Hubble Legacy Archive at {https://hla.stsci.edu}.
The Gemini spectroscopic data from GMOS-N are available from {https://archive.gemini.edu}. The Keck Observatory data
used here will be available in September 2021 from {https://www2.keck.hawaii.edu/koa/public/koa.php} and are available prior to that
on reasonable request from the corresponding author. The software package ptmatch is available at {https://github.com/ntessore/imagemap.

% The best way to enter references is to use BibTeX:
%% \bibliographystyle{plain}
%% \bibliography{hamilton}

%% \section{Appendices}
%%% \appendix

%\section{Table 1: Nuclear Bulge Magnitudes}

% Don't change these lines
\bsp	% typesetting comment

\label{lastpage}

\end{document}